\documentclass[a4paper,fleqn,usenatbib]{mnras}

\usepackage{newtxtext,newtxmath}
\usepackage[T1]{fontenc}
\usepackage{ae,aecompl}

\usepackage{graphicx}	% Including figure files
\usepackage{amsmath}	% Advanced maths commands
\usepackage{amssymb}	% Extra maths symbols

\usepackage{amssymb}
\usepackage{lipsum}
\usepackage{afterpage}
\usepackage{threeparttable}
\usepackage{placeins}
\usepackage{multirow,makecell,booktabs}
\usepackage{subfigure}
\usepackage[nameinlink]{cleveref}
\Crefname{figure}{Fig.}{Figs.}
\Crefname{equation}{Eq.}{Eqs.}
\usepackage{enumitem}
\usepackage[bigfiles]{media9}
\newcommand{\hi}{\mbox{H{\sc i}}}
\newcommand{\kms}{km s$^{-1}$}

\newcommand{\Mo}{\rm M_{\odot}}

\newcommand{\dg}{^{\circ}}
\newcommand{\mjybeam}{\rm mJy\,beam^{-1}}

\newcommand{\cm}{\rm cm^{-2}}

\newcommand{\ddeg}{\overset{\circ}{.\,}}
\newcommand{\dmin}{\overset{\prime}{.}}

\newcommand{\e}[1]{\times 10^{#1}}
\newcommand{\HI}{\textsc{Hi}}

\newcommand{\miriad}{\textsc{Miriad}}
\newcommand{\gipsy}{\textsc{Gipsy}}

\newcommand{\moment}{\textsc{moment}}

\newcolumntype{H}{>{\setbox0=\hbox\bgroup}c<{\egroup}@{}} %%% Hide column
\title[Deep \HI\ view of M81]{A $5\dg\times5\dg$ deep \hi\ survey of the M81 group}

\author[Sorgho et al.]{
A. Sorgho$^{1}$\thanks{sorgho@ast.uct.ac.za},
T. Foster$^{2}$,
C. Carignan$^{1,3}$,
L. Chemin$^{4}$
\\
\\
$^{1}$ Department of Astronomy, University of Cape Town, Private Bag X3, Rondebosch 7701, South Africa\\
$^{2}$ Dominion Radio Astrophysical Observatory, P.O. Box 248, Penticton, British Columbia, V2A 6J9, Canada\\
$^{3}$ Laboratoire de Physique et de Chimie de l’Environnement, Observatoire d’Astrophysique de l’Universit\'{e} Ouaga I Pr Joseph Ki-Zerbo (ODAUO),\\ \,\,\, 03 BP 7021, Ouaga 03, Burkina Faso\\
$^{4}$ Centro de Astronom\'{i}a (CITEVA), Universidad de Antofagasta, Avenida Angamos 601, Antofagasta, Chile
}

\date{Accepted XXX. Received YYY; in original form ZZZ}

\pubyear{2017}

\begin{document}
\label{firstpage}
\pagerange{\pageref{firstpage}--\pageref{lastpage}}
\maketitle

\begin{abstract}
%We present new DRAO \hi\ observations of the region of the M81 group covering the M81 system, NGC 2976 and IC 2574.
A $25\,\rm deg^2$ region, including the M81 complex (M81, M82, NGC 3077), NGC 2976 and IC2574, was mapped during $\sim3000$ hours with the DRAO synthesis telescope.
With a physical resolution of $\sim1$ kpc, these observations allow us to probe a large region down to column density levels of $\sim1\e{18}\,\cm$ over 16 \kms, mapping the extent of the \hi\ arm connecting the system and NGC 2976, and resolving the \hi\ clouds adjacent to the arm. The observations also reveal a few clouds located between the system and IC 2574, probably tidally stripped from a past interaction between the two systems. Given the regular velocity distribution in the \hi\ envelope of the system, we attempt and derive an idealised large-scale rotation curve of the system.
%which shows the effects of the companion galaxies (mainly M82) on the rotation of the system using the tilted-ring model.
We observe a flat trend for the rotation velocity of the overall system from 20 kpc out to 80 kpc, well beyond the outskirts of the M81 disk, although with asymmetries like a wiggle at the vicinity of M82. This supports the assumption that intergalactic gas and galaxies in the system participate to a large-scale ordered rotation motion which is dominated by M81.
%Although the rotation curves of the approaching and receding sides of the system tend to differ, we observe a flat trend in the rotation of the overall system that corroborates the assumption of a large-scale rotating system dominated by the M81 galaxy.
Also, our \hi\ analysis of the group further supports the hypothesis that the galaxies forming the system moved closer from afar, in agreement with numerical simulations.
\end{abstract}


%%%%%%%%%%%%%%%%%%%%%%%%%%%%%%%%%%%%%%%%%%%%%%%%%%%%%%%%%%%%%%%%%%%%%
%%%%%%%%%%%%%%%%%%%%%%%% INTRODUCTION %%%%%%%%%%%%%%%%%%%%%%%%%%%%%%%

\section{Introduction}
Galaxy interactions are a key process driving the evolution of galaxies in the Universe. The theory predicts that the morphology and gas content of galaxies are shaped by the interactions with their environment and counterparts \citep[e.g.,][]{Toomre1972,Gunn1972,Dressler1980}. In particular, tidal interactions are known to affect the stellar disk and star formation rates of galaxies \citep[e.g.,][]{Cayatte1990,Hernquist1992,Mihos1996,VanDokkum1999,Hopkins2006}. However, the component of galaxies through which the impact of galaxy interactions is most easily seen is the neutral hydrogen (\hi) envelope, which is known to often extend farther than the stellar disk. In fact, observations of several nearby galaxies exhibiting \hi\, envelopes revealed signs of interactions with either their environment \citep[e.g.,][]{Vollmer2008b,Kenney2014}, or with other galaxies \citep[e.g.,][]{Hibbard1996}. One of the best examples of these galaxies is arguably the M81 group, whose main galaxy members (M81, M82 and NGC 3077) seem to share a common \hi\, envelope although their optical disks are distinct.

%%% ENVIRONMENT??? The HI is known to extend farther than the optical disk

%%% The M81 group
The M81 group has been the subject of many \hi\, observational studies over the last two decades. However, most of these studies either focused on individual galaxies in the group at high angular resolution \citep[e.g. THINGS;][]{Walter2008}, or covered a large field of view but lacked angular resolution \citep[e.g. GBT observations,][]{Chynoweth2008,Chynoweth2009}. The first high resolution ($1'$), wide field ($1.7\dg\times1.6\dg$) map of the group was obtained by \citet{Yun1994} with the VLA telescope, and revealed \hi\, complexes in the system down to \hi\, mass limits of $10^6\,\Mo$. However, the velocity resolution (10.3 \kms) and \hi\, column density sensitivity ($\sim2\e{20}\,\cm$) of the observations conducted by the authors are limited, compared to the above-mentioned THINGS and GBT observations. Recently, a higher resolution VLA map\footnote{VLA data publicly available at \url{http://www.astron.nl/~blok/M81data/}} of the system was presented by \citet{DeBlok2018a}, where the discovery of a number of kpc-sized low-mass \hi\, clouds in the vicinity of the M81 complex was reported. The authors achieved a high column density sensitivity level, with a mass detection limit of a few times $10^4\,\Mo$; but like in \citet{Yun1994} and the GBT observations, the area covered in these observations only includes the complex and the dwarf galaxy NGC 2976 located south-west of the complex.
The map presented by \citet{Yun1994} revealed a filamentary structure located west of M81 (but not connected to the complex), and spanning a length of only $\sim21$ kpc. In the subsequent high sensitivity map of \citet{Chynoweth2008} with the GBT, the filament appeared to connect the complex to NGC 2976. However, due to the limited resolution of the GBT ($\sim10$ kpc), it was impossible to resolve the individual clouds and as a result, it was not clear whether the apparent filament was made of discrete clouds or of a long single tail connecting the two structures. The recent observations of \citet{DeBlok2018a} provided a partial answer to these questions, showing that the filament is in fact a series of discrete clouds connected to the M81 system, but not to NGC 2976. However, the lowest contour in the authors' total \hi\ intensity map is $3\e{19}\,\cm$. It is quite likely that an increased sensitivity could allow the detection of a more extended \hi\, filament, possibly connecting to NGC 2976.

To date, none of the existing deep \hi\ observations has ever covered an area of the group wide enough to include both the M81 system and the distant eastern member of the group, IC 2574. Located at a projected distance of 193 kpc from the centre of M81 group, the galaxy IC 2574 has a systemic velocity coinciding with that of the M81 system, which suggests a possible kinematical association with the system. Moreover, the existence  of an ``optically dark'' galaxy -- HIJASS J1021+68 -- located west of IC 2574 has been reported in the recent blind \hi\ HIJASS survey \citep[\hi\ Jodrell All-Sky Survey,][]{Boyce2001}. This hints at the existence of more low density clouds in the space between the M81 system and IC 2574, and a possible \hi\ connection of the two entities through discrete clouds. A detection of such clouds, which is only achievable through deep \hi\ observations, could offer new insights into the dynamics of the galaxy members of the M81 system. 

The present observations aim, primarily, at not only making a full census of the \hi\ clouds around the M81 system and in the region extending towards IC 2574, but also mapping at a high resolution the full extent of the \hi\ streamers, filaments and tails connecting the main group members. We use the Synthesis Telescope (ST) of the Dominion Radio Astrophysical Observatory\footnote{\url{https://www.nrc-cnrc.gc.ca/eng/solutions/facilities/drao.html}} (DRAO) to map, at high resolution and column density sensitivity, the \hi\, in and around the M81 complex, including the area located East of the complex and extending out to include the galaxy IC 2574. A total of 30 fields were observed (including archival data), covering an area of 25 squared degrees at a spatial resolution of $\sim1'$, which, at the distance of M81 \citep[3.63 Mpc,][]{Karachentsev2004}, corresponds to $1.05$ kpc. This resolution is $\sim2.5$ times lower than that of \citet{DeBlok2018a}, but the achieved sensitivities are sensibly similar.  

To achieve an even higher sensitivity than the previous VLA observations, we smooth the DRAO observations to a lower spatial resolution, in an attempt to not only extend the filament between the system and NGC 2976, but also detect any previously undetected gas cloud lying in the vicinity of M81. In fact, numerous high velocity clouds have been detected around M81 (and in the NGC 2403 group) in a GBT observational campaign of the region conducted by \citet{Chynoweth2008}, \citet{Chynoweth2009} and \citet{Chynoweth2011}, down to an \hi\, mass limit of a few times $10^6\,\Mo$. However, when compared to cosmological N-body simulations, \citet{Chynoweth2011} found that the number of detected \hi\, clouds is significantly lower than predicted by the simulations. Moreover, the authors found a mismatch between the simulated minihalos and the phase space of the observed clouds. The implication of these results is that there must exist a vast population of smaller (<10 kpc) \hi\, clouds not resolved by the GBT, likely generated by the galaxies themselves and through tidal stripping caused by galaxy interactions. This is confirmed by the \citet{DeBlok2018a}'s observations where the authors reported the detection of newly discovered kpc-sized low-mass \hi\, clouds with masses of a few times $10^6\,\Mo$. This constitutes an evidence that increasing the sensitivity of the data can lead to an increase of the number of new detections.

This paper, which is the first of a series, presents the global results of the deep \hi\ survey of the M81 group. We present here the distribution and kinematics of the \hi\ gas in the group, and present global \hi\ properties of the main group members. In forthcoming papers, we will investigate more deeply the local properties of the \hi, especially the \hi\ in the discrete clouds, and study the connections between the galaxies. The present paper is organised as follows: in \Cref{sec:obs-red} we describe the observations conducted with the DRAO ST and the data reduction method. Next, we present the \hi\ profiles of the M81 system member galaxies and the \hi\ maps obtained from the observations in \Cref{sec:hi-properties} where we discuss the new features. A kinematical study of the M81 group is presented in \Cref{sec:kinematics}, and we discuss the possible dynamical evolutions of the main galaxies of the group in \Cref{sec:discussions}. Finally, we summarise our results and give brief conclusions in \Cref{sec:conclusions}.

%%%%%%%%%%%%%%%%%%%%%%%%%%%%%%%%%%%%%%%%%%%%%%%%%%%%%%%%%%%%%%%%%%%%%

%%%%%%%%%%%%%%%%%%%%%%%%%%%%%%%%%%%%%%%%%%%%%%%%%%%%%%%%%%%%%%%%%%%%%
%%%%%%%%%%%%% OBSERVATION & DATA REDUCTION %%%%%%%%%%%%%%%%%%%%%%%%%%

\section{Observations and data reduction}\label{sec:obs-red}

\subsection{Observations}\label{sec:obs}
The observations were conducted between 2016 and 2018 using the DRAO telescope. The goals of the observations are as follow: i) detect and better resolve \hi\ streamers, filaments and tails that connect group members over a larger area of the group than the existing \hi\ maps, ii) follow up on the 9 \hi\, clouds discovered in the GBT studies \citep{Chynoweth2008,Chynoweth2011} to better resolve them and clarify their relationship with the galaxies and intergalactic \hi, and iii) add to the number of newly detected \hi\, clouds in the M81 group, and search among these for possible candidate \hi\, clouds associated with dark matter minihaloes, or new clouds analogous to the enigmatic population in the M31-M33 system \citep{Wolfe2013,Wolfe2016}.

\begin{figure*}
\makebox[\textwidth][c]{
\hspace{30pt}
\includegraphics[width=1.1\textwidth]{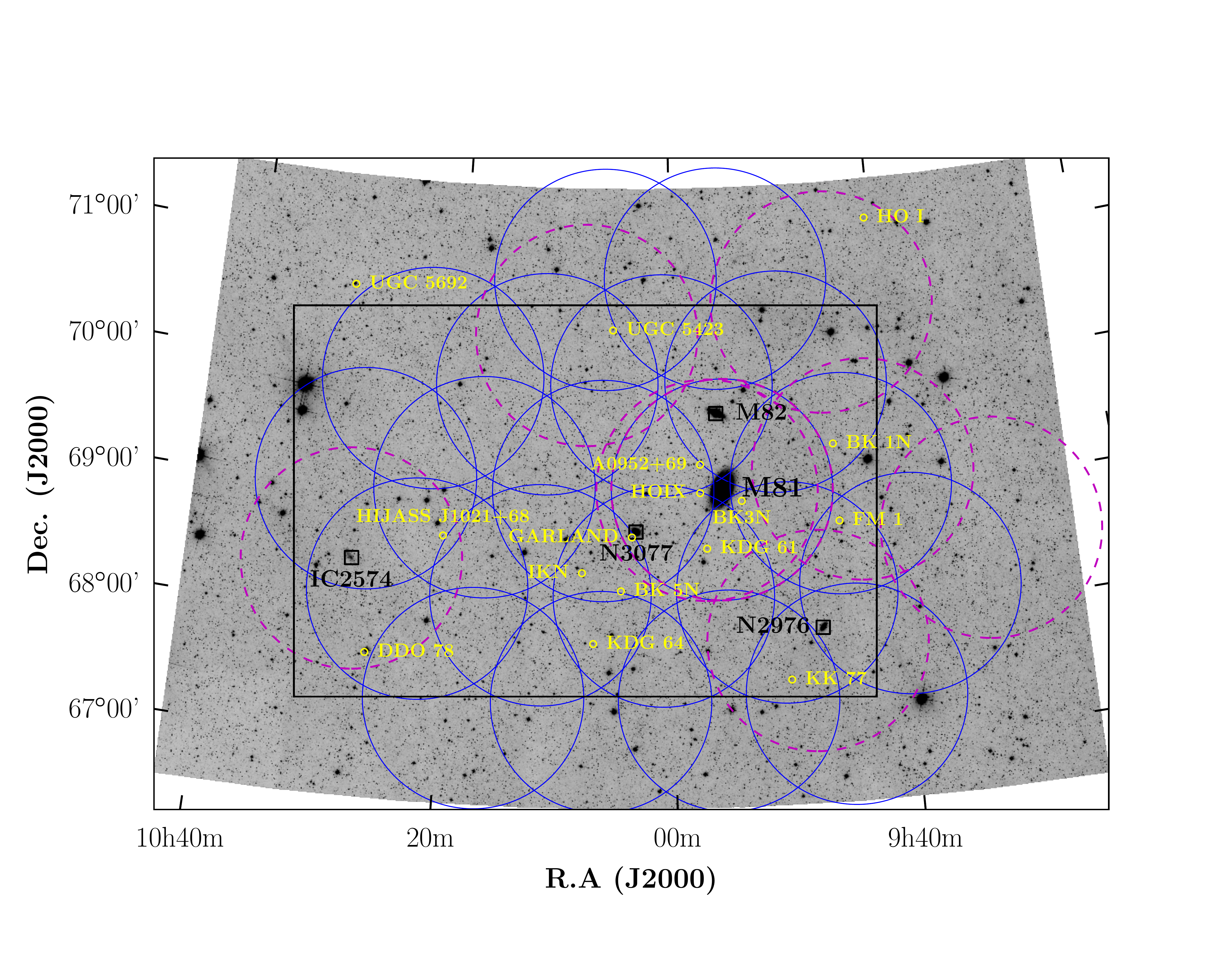}}
\vspace{-40pt}
\caption{The mosaic fields observed with the DRAO ST overlaid on a {\it WISE} W1 grayscale image of the region. The {\it blue circles} show the original fields while the {\it magenta dashed circles} represent the archived data. The rectangle shows the area of the moment maps presented in \Cref{fig:nomw,fig:nhimap_drao,fig:vfield_drao,fig:vdisp_drao}. The major galaxies of the group are labelled in {\it black squares} while the dwarf galaxies are labelled in {\it yellow circles}.}\label{fig:fields}
\end{figure*}

With a primary beam full-width at half-power of $107\dmin2$ in the 21cm, the DRAO ST consists of 7 antennas of 9m diameter each, arranged such that they form an east-west baseline with a maximum separation of 617.1m \citep[see][for telescope specifications]{Landecker2000}. The telescope was used to observe, at 1420 MHz, a total of 20 fields (for 144 hrs each) centred on a 25 $\rm deg^2$ region including the M81 complex, IC 2574 and NGC 2976. Moreover, 10 additional fields (with four centred on two different coordinates) were recovered from the DRAO archive and added to the observed data to increase the SNR in certain key areas (around M81, IC 2574, NGC 2976, and the area between M81 and NGC 2976). In \Cref{tb:fields} we give the centre coordinates of each of these fields, which are overlaid on an optical {\it WISE} \citep{Wright2010} W1  grayscale image of the region in \Cref{fig:fields}. With a spacing of $\Delta\sim 57'$, the centres of the 20 individual observed fields were chosen to give a uniform sensitivity across the region containing the major galaxy members of the group, while allowing to look for \hi\, in clouds and filaments connecting and bridging galaxy members. It also allows to search for more widespread massive \hi\, clouds associated with the DM halos of the group, whose existence is predicted by numerical simulations and expected to lie within 1 Mpc of the major DM haloes \citep{Blitz1999}. Upcoming observations with the DRAO ST will increase the sensitivity in the eastern region of the mosaic, around the galaxy IC 2574.

The spectrometer used for the observations provides a 4 MHz bandwidth distributed across 256 channels, which corresponds to a channel width of 3.3 \kms\, for a spectral resolution of about 5.3 \kms. The observations were centred on the velocity $-36.7$ \kms, which is approximately the central velocity of the complex, and cover the range $-459$ \kms\, to 382 \kms. 
The working observed noise at the field centres was on average 1.35 K per channel (or $\sim6\,\mjybeam$ per channel) and varied only by 0.07 K from field to field.
The archived fields are all centred on 0 \kms\, velocity and although they do not capture the entire velocity width of \hi\ from the M81 system, after reprocessing to the same spatial and velocity resolution as the 20 new fields, they have similarly low rms noise levels in the range of $1.3-1.4$ K per channel. Six of these fields (all except HP03, JG5, JG6 and MB) have a bandwidth of 1 MHz and originally spanned only 64 channels, ranging in velocity from about $-105$ \kms\ to 105 \kms.
The field HP03, like the present observations, has a bandwidth of 4 MHz, covering the range $-422$ \kms\ to 422 \kms. JG5 and JG6 were observed using the 2 MHz bandwidth, making their velocity coverage $-211$ to 211 \kms. As for the field MB, the bandwidth is 0.5 MHz and also centred on 0 \kms, which corresponds to the narrower range of about $-53$ \kms\, to 53 \kms.

As an interferometer, the ST is limited in its ability to accurately recover the fluxes of large scale structures, i.e those at angular sizes larger than $56'$, as given by the telescope's shortest baseline of 12.9m. To compensate for this, single dish observations are combined to interferometric observations to perform a short-spacing correction. In this work, we used single-dish data from the 100m Effelsberg telescope obtained as part of the EBHIS \hi\, survey \citep{Winkel2016} to apply this short-spacing correction to the ST observations. The EBHIS data have a spatial resolution of $10\dmin8$ and a full spectral resolution of 1.44 \kms, and covers our whole field both spatially and in velocity.

\begin{table*}
	\centering
	\begin{tabular}{l c c c c c c c c c c}
	\hline
    \hline
	Field ID & \multicolumn{2}{c}{J2000 position} & Obs. date & $\theta_{\rm maj}\times\theta_{\rm min}$ & $\theta_{\rm PA}$ & $\Delta B_{\rm tot}$ & $\Delta v$ & $\Delta v_{\rm res}$ & $v$ range & rms\\
	& R.A & Dec. & (YYYY/MM) & ($''$) & ($\dg$) & (MHz) & (\kms) & (\kms) & (\kms) & ($\mjybeam$)\\
	(1) & \multicolumn{2}{c}{(2)} & (3) & (4) & (5) & (6) & (7) & (8) & (9) & (10)\\
	\hline
	MG1  & 09:55:33.17 & +69:03:55.0 & 2015/09 & $61.6\times59.0$ & -90.8 & 4.0 & 3.30 & 5.28 & -459 -- 382 & 8.40\\
	MG2  & 09:44:48.11 & +69:03:55.0 & 2015/09 & $62.1\times59.4$ & -81.4 & 4.0 & 3.30 & 5.28 & -459 -- 382 & 8.24\\
	MG3  & 09:50:10.64 & +69:55:06.6 & 2015/10 & $60.8\times59.1$ & -91.9 & 4.0 & 3.30 & 5.28 & -459 -- 382 & 7.90\\
	MG4  & 10:00:55.70 & +69:55:06.6 & 2015/10 & $61.1\times59.2$ & -86.3 & 4.0 & 3.30 & 5.28 & -459 -- 382 & 7.77\\
	MG5  & 10:06:18.23 & +69:03:55.0 & 2016/01 & $60.8\times59.5$ & -101.4 & 4.0 & 3.30 & 5.28 & -459 -- 382 & 8.22\\
	MG6  & 10:00:55.70 & +68:12:43.4 & 2016/01 & $61.9\times59.2$ & -94.4 & 4.0 & 3.30 & 5.28 & -459 -- 382 & 7.99\\
	MG7  & 09:50:10.64 & +68:12:43.4 & 2016/01 & $61.3\times59.8$ & -90.2 & 4.0 & 3.30 & 5.28 & -459 -- 382 & 8.23\\
	MG8  & 10:11:40.76 & +69:55:06.6 & 2015/11 & $61.1\times59.2$ & -94.0 & 4.0 & 3.30 & 5.28 & -459 -- 382 & 8.12\\
	MG9  & 09:39:25.58 & +68:12:43.4 & 2016/06 & $61.1\times60.0$ & -74.6 & 4.0 & 3.30 & 5.28 & -459 -- 382 & 8.52\\
	MG10 & 10:06:18.23 & +70:46:18.2 & 2016/06 & $61.6\times58.9$ & -76.8 & 4.0 & 3.30 & 5.28 & -459 -- 382 & 8.14\\
	MG11 & 09:44:48.11 & +67:21:31.8 & 2016/06 & $62.2\times59.4$ & -81.9 & 4.0 & 3.30 & 5.28 & -459 -- 382 & 8.56\\
	MG12 & 09:55:33.17 & +70:46:18.2 & 2016/06 & $60.7\times58.8$ & -84.1 & 4.0 & 3.30 & 5.28 & -459 -- 382 & 8.26\\
	MG13 & 09:55:33.17 & +67:21:31.8 & 2016/08 & $62.4\times59.2$ & -89.5 & 4.0 & 3.30 & 5.28 & -459 -- 382 & 8.05\\
	MG14 & 10:06:18.23 & +67:21:31.8 & 2016/08 & $62.4\times59.2$ & -88.8 & 4.0 & 3.30 & 5.28 & -459 -- 382 & 8.43\\
	MG15 & 10:11:40.76 & +68:12:43.4 & 2016/08 & $62.1\times59.0$ & -87.1 & 4.0 & 3.30 & 5.28 & -459 -- 382 & 8.12\\
	MG16 & 10:17:03.29 & +69:03:55.0 & 2016/09 & $61.5\times59.1$ & -81.7 & 4.0 & 3.30 & 5.28 & -459 -- 382 & 8.43\\
	MG17 & 10:22:25.82 & +69:55:06.6 & 2016/10 & $61.3\times58.9$ & -84.1 & 4.0 & 3.30 & 5.28 & -459 -- 382 & 8.28\\
	MG18 & 10:17:03.29 & +67:21:31.8 & 2016/10 & $62.7\times58.9$ & -88.6 & 4.0 & 3.30 & 5.28 & -459 -- 382 & 8.41\\
	MG19 & 10:22:25.82 & +68:12:43.4 & 2016/11 & $62.1\times59.2$ & -83.9 & 4.0 & 3.30 & 5.28 & -459 -- 382 & 8.09\\
	MG20 & 10:27:48.35 & +69:03:55.0 & 2017/01 & $61.5\times58.8$ & -68.0 & 4.0 & 3.30 & 5.28 & -459 -- 382 & 7.70\\
	HP03$^*$& 09:55:33.17 & +69:03:55.0 & 2014/03 & $60.0\times59.6$ & -79.1 & 4.0 & 3.30 & 5.28 & -422 -- 422 & 7.48\\
	JG5$^*$ & 09:55:33.17 & +69:03:55.0 & 2012/07 & $62.7\times58.2$ & -74.5 & 2.0 & 1.65 & 2.64 & -211 -- 211 & 8.80\\
	JG6$^*$ & 10:28:23.48 & +68:24:44.0 & 2012/07 & $61.6\times59.1$ & -94.7 & 2.0 & 1.65 & 2.64 & -211 -- 211 & 8.32\\
	NN$^*$  & 09:45:20.69 & +70:32:11.3 & 2009/05 & $61.2\times58.5$ & -92.7 & 1.0 & 0.84 & 1.32 & -105 -- 105 & 8.01\\
	PO$^*$  & 10:08:02.65 & +70:19:16.1 & 2010/12 & $62.1\times59.2$ & -102.5 & 1.0 & 0.84 & 1.32 & -105 -- 105 & 8.01\\
	PP$^*$  & 09:47:46.22 & +67:48:35.4 & 2010/11 & $63.8\times58.7$ & -92.1 & 1.0 & 0.84 & 1.32 & -105 -- 105 & 8.91\\
	MB$^*$  & 09:56:54.88 & +69:03:50.5 & 2005/07 & $62.7\times58.2$ & -67.2 & 0.5 & 0.41 & 0.66 & -53 -- 53 & 8.64\\
	MB2$^*$ & 09:56:54.88 & +69:03:50.5 & 2009/01 & $61.3\times59.0$ & -97.1 & 1.0 & 0.84 & 1.32 & -105 -- 105 & 8.00\\
	PM$^*$  & 09:42:42.72 & +69:09:55.2 & 2011/03 & $61.3\times58.8$ & -87.1 & 1.0 & 0.84 & 1.32 & -105 -- 105 & 8.43\\
	PN$^*$  & 09:31:48.72 & +68:35:05.2 & 2011/03 & $62.1\times58.7$ & -91.5 & 1.0 & 0.84 & 1.32 & -105 -- 105 & 8.47\\
	\hline
	\end{tabular}
	\caption{The parameters of the individual fields of the DRAO mosaic. The archived fields are marked with an asterisk. Column (1): Field ID; Column (2): J2000 coordinates of the field centre; Column (3): Date of observation; Column (4): Beam size; Column (5): Position angle of the beam; Column (6): Total bandwidth of the spectrometer used for observation; Column (7): Channel spacing of the corresponding cube; Column (8): Velocity resolution; Column (9): Range of velocity covered; Column (10): Noise per channel in the cube.}\label{tb:fields}
\end{table*}

\subsection{Data reduction}\label{sec:reduction}
%The data reduction technique used is similar to that described in \Citet{Chemin2009}. It consists of processing the individual datacubes obtained from the respective fields before combining them into a mosaic. The first step of the reduction is a continuum subtraction of the datacubes. 
Calibration, reduction and data-processing on individual datacubes for each of the fields are undertaken before combining them into a mosaic. The approach used is essentially identical to that developed for the Canadian Galactic Plane Survey \citep[see Section 2.2,][]{Taylor2003}, and also used in \Citet{Chemin2009}. \hi\ 21cm line (spectrometer) DRAO ST data are bandpass calibrated immediately post-observation, after correlated UV data are formed. At this stage continuum data are brightness-temperature (total intensity) calibrated through observations of sources 3C48, 3C147, 3C286 and 3C295. The spectrometer data are not initially total-intensity calibrated with these astronomical sources at the same time as the continuum; rather this is done at a later stage. Instead, we first perform a channel-by-channel subtraction of the continuum emission in the datacubes. This is achieved by averaging line-free channels at both ends of each datacube, providing two continuum maps (respectively at the low and high velocity ends) that are then linearly interpolated in the velocity space to obtain the continuum level. This continuum level is then subtracted from each channel map of the datacube, providing a continuum-subtracted datacube that is then calibrated. DRAO field observations are bracketed by observing a set of strong compact and non-time variable calibrator sources \citep[3C48, 3C147, 3C286 and 3C295; see][]{Landecker2000}, and continuum maps in each of the 1420 MHz and 408 MHz bands are flux-density calibrated and phase referenced in the post-processing stages of the data product preparation. We use these calibrated continuum maps to calibrate the \hi\ line data as discussed below.
The second step consists in performing the calibration, for which two continuum maps are created for each individual datacube: one created by CLEANing (around strong sources) a calibrated continuum map of each field obtained in the 30 MHz-wide band of the DRAO, and a second one created by averaging the above-mentioned end-channels continuum maps. A comparison of the fluxes of point sources between these two continuum maps is then performed, allowing to calibrate the fluxes in the cubes. This calibration method provides uncertainties of about 5\%. At this point, the individual fields are continuum-subtracted and calibrated, and spatially overlap as shown in \Cref{fig:fields}. They are then added together following an rms-weighting scheme, to form a single mosaic datacube of $1024\times1024\times256$ pixels in size, where each pixel is $20''$ wide.

The final step of the reduction consists in adding the single-dish EBHIS data to the DRAO ST data to apply the short-spacing correction. This was done in the image plane, and consisted of smoothing the ST mosaic to the same resolution as the EBHIS data ($10\dmin8$) then subtracting the smoothed mosaic from the EBHIS data to leave the missing structures that the DRAO ST did not detect. The difference cube, essentially containing structures invisible to the ST, was then added back to the full resolution DRAO mosaic. Due to the careful calibration of our interferometer data and the excellent overlap between spatial structures observed by the ST and the Effelsberg telescopes, the final short-spacings corrected DRAO mosaic shows no negative bowls around structures as large as the MW \hi, and as small as the M81/82 galaxies themselves. 

\begin{figure}
\makebox[\columnwidth][c]{
\hspace{-20pt}
\includegraphics[width=1.1\columnwidth]{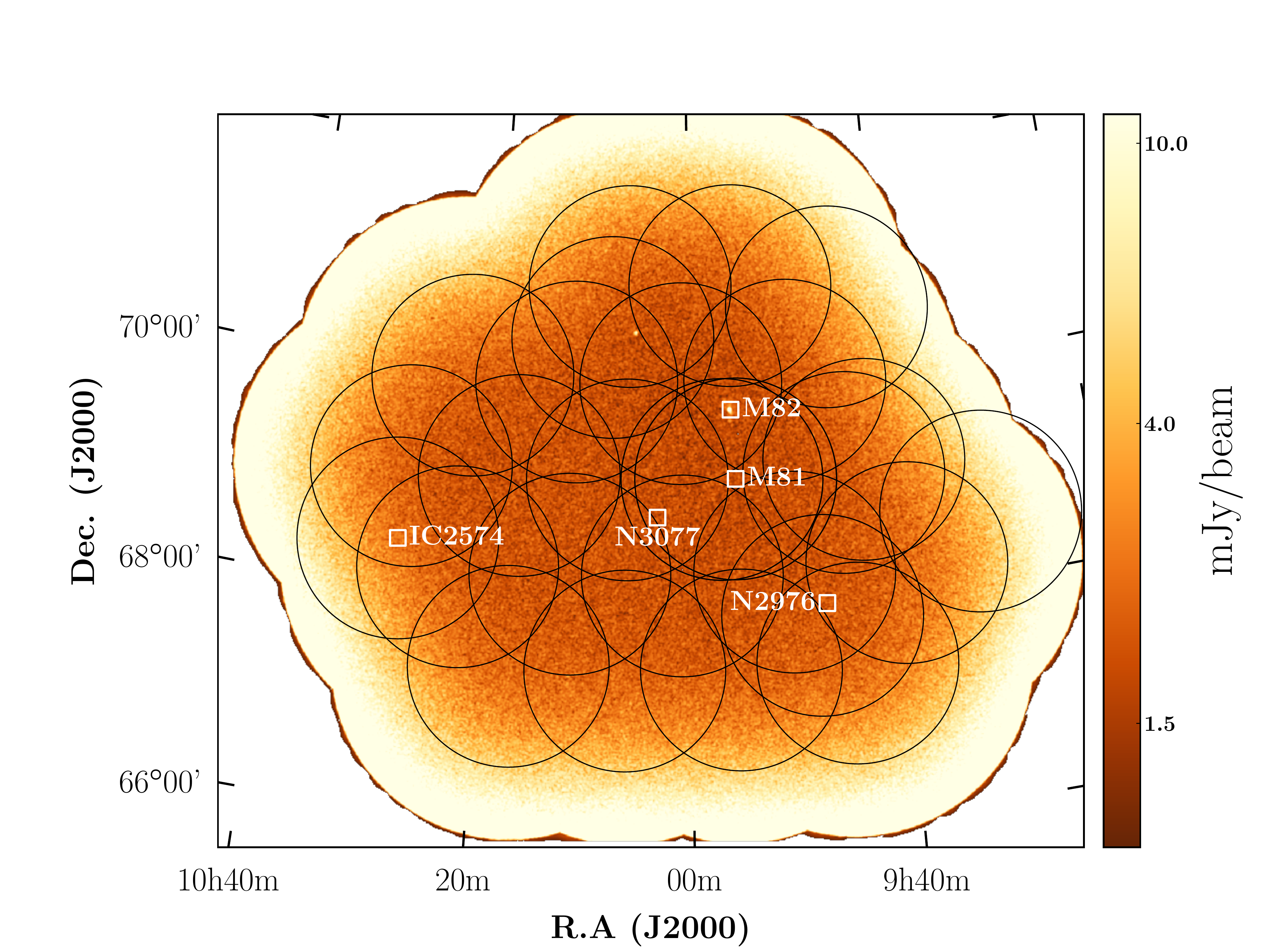}}
\vspace{-15pt}
\caption{Sensitivity distribution map of the mosaic, derived from the line-free channels of the data where the noise is uniform over the central region. The black circles show the individual fields of \Cref{fig:fields}.}\label{fig:sensmap}
\end{figure}

The spectral and spatial resolutions of the short-spacing corrected datacube are 5.2 \kms\, and $61''\times59''$ respectively, with a channel width of 3.3 \kms\ and a pixel size of $20''$. The average noise around the M81/82 system is $3.6\,\mjybeam$, increasing to $4.8\,\mjybeam$ around NGC 2976, and reaching $5.3\,\mjybeam$ around IC 2574. In terms of column densities at a 16 \kms\ velocity width, these noise levels correspond to $1.8\e{19}\,\cm$, $2.3\e{19}\,\cm$ and $2.6\e{19}\,\cm$ respectively around M81, NGC 2976 and IC 2574. In \Cref{fig:sensmap} we show a sensitivity map of the datacube, derived from the emission-free channels in the cube. The noise values in the map were determined from a total of 41 emission-free channels (out of a total of 244 channels\footnote{Twelve of the 256 channels, located at the edges of the cube, were removed because of strong artefacts.}) selected at each spectral end of the datacube. We also overplot in the figure the different DRAO ST fields used for the mosaic.

To increase the column density sensitivity of the observations, we have also produced a smoothed version of the cube, with a lower angular resolution of $1.8'$. At the distance of M81, this translates to a physical resolution of 1.9 kpc. 
This low resolution cube has a spatial resolution of $108''\times108''$, and a velocity resolution of 5.2 \kms. Like the full-resolution cube, it also has a channel width of 3.3 \kms. The average noise in the cube is $5.5\,\mjybeam$ around the M81 system, which translates to a $1\sigma$ column density sensitivity of $1.1\e{18}\,\cm$ at a velocity width of 16 \kms. Because of its higher sensitivity, we will only use this smoothed version of the cube for the rest of the analysis in this paper.

\subsection{Milky Way emission}\label{sec:mwremoval}
A major problem with mapping the \hi\, in galaxies at low systemic velocity is the contamination from the Milky Way (MW). Several techniques exist to remove the MW's foreground \hi\, from observations, ranging from manual mitigations to statistical methods \citep[e.g.][]{Sofue1979}. \citet{Chynoweth2008} performed, in their work, a spectral interpolation of the signal-of-interest over the contaminated channels, allowing to ``guess'' the shape of the target emission in these channels. Although this technique provides an acceptable representation of the targeted emission, it does not provide an accurate description of the global profile of the emission, especially for high spectral (and spatial) resolution.

\begin{figure}
\makebox[\columnwidth][c]{
\includegraphics[width=1.15\columnwidth]{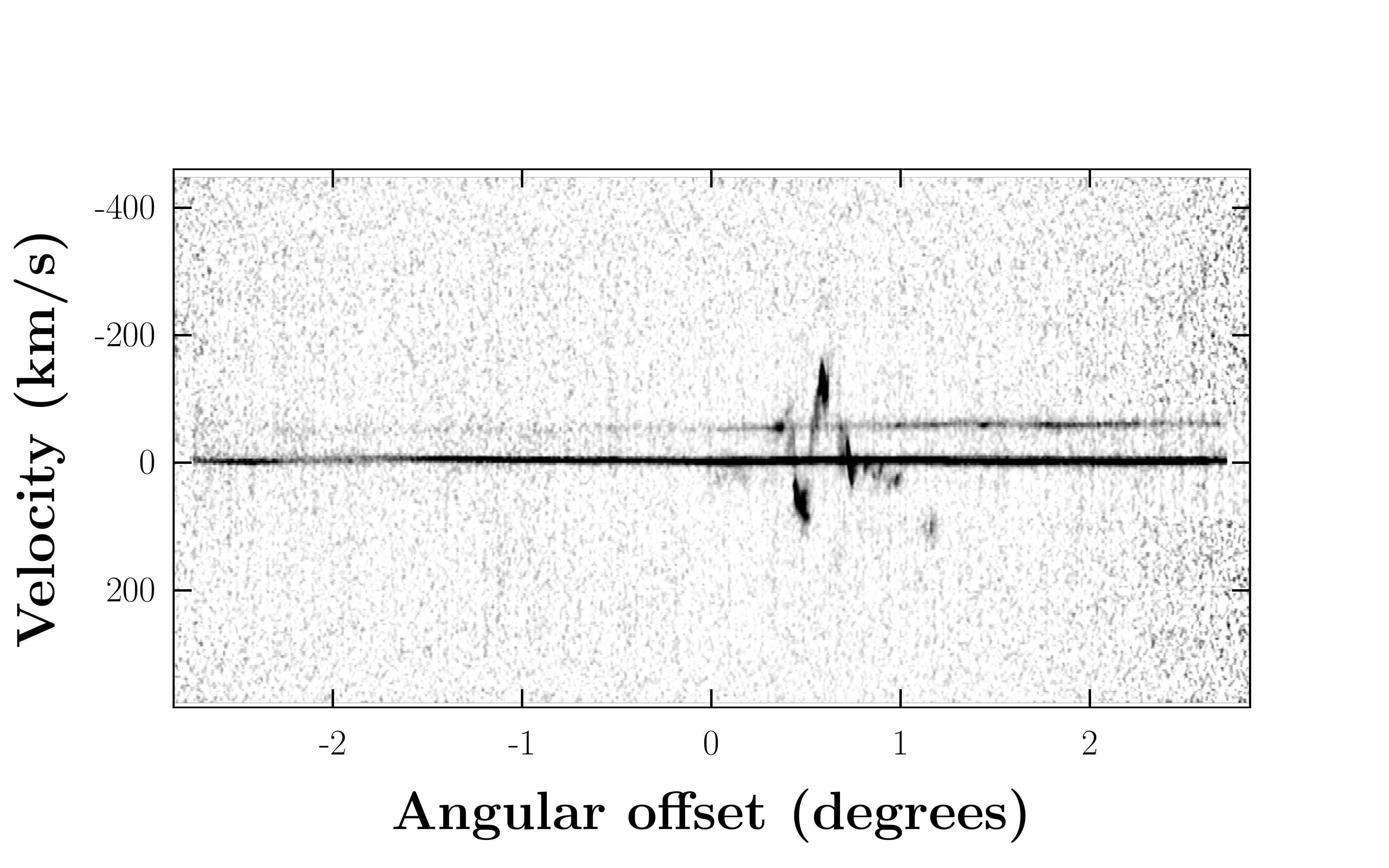}}
\vspace{-15pt}
\caption{Position-velocity slice of the DRAO cube, showing the contamination of the data by the foreground MW emission. The two horizontal lines (centred at $-56.5$ \kms\, and $-2.1$ \kms) show the peaks of the Galactic \hi.}\label{fig:pvslice}
\end{figure}

The technique used in the present work is somewhat similar to that described in \citet{Chemin2009}, and consists of subtracting a modelled MW emission from the observations. We started by identifying the channels affected by the MW emission and found a total of 24 channels, covering the velocity range $-76.3$ \kms\ to $-40.0$ \kms\ and $-23.5$ \kms\ to $19.3$ \kms. In \Cref{fig:pvslice} we present a position-velocity diagram obtained by considering a horizontal slice cutting through the centre of M81, which illustrates the two peaks of the MW emission at $-56.5$ \kms\ and $-2.1$ \kms, respectively.
Next, we constructed a total \hi\ intensity (zeroth moment) map from the MW-free channels, based on a mask created using the ``smooth and clip'' method described in \Cref{sec:profs_moms} below. This MW channels-free intensity map, presented in \Cref{fig:nomw}, was obtained by blanking the above-mentioned 24 channels affected by MW emission.
The map was then used as a mask to identify and blank the emission of the M81 system from the data. This means that any pixel in the contaminated channels that has a non-zero value in the constructed moment map was blanked. This leaves us with a datacube which has no M81 emission in the MW-contaminated channels. In a subsequent step, we used a two-dimensional gaussian kernel of standard deviation equals to $4.2'$ (equivalent to about 2.3 beams) to interpolate from the non-blanked pixels and replace the previously blanked pixels that presumably represent the emission from the M81 system. Kernels of smaller sizes were found to poorly model the MW emission and therefore produce artefacts in the inner regions of the M81 system, and larger sizes sensibly provide similar results. 
The interpolation is done individually for all the planes that are MW-contaminated and, in theory, fills the M81 pixels (i.e, pixels that belong to the M81 system) with Galactic foreground emission. Given the high level of foreground contamination of the data, it is best to rule all emission in the affected channels and outside the M81 system as noise or galactic foreground emission. Therefore, the pixels in the affected channels that lie outside M81 (i.e, the pixels outside the moment map) can be discarded by either setting them to zero, or to a pure statistical noise value. In practice, the second option is equivalent to replacing these pixels with their counterparts from emission-free channels selected outside the velocity range of M81. This produces a datacube in which the MW emission is almost completely removed.

An inspection of the resulting cube showed that the procedure described above successfully mitigated most of the foreground \hi\ in the data. However, because of the zero spacing correction of the data, the entire planes of a few channels (where the foreground emission peaks) were completely covered with bright MW emission. No emission from the M81 system could be detected in these channels because of their low flux relative to the Galactic \hi. The applied technique did not succeed at completely removing the foreground emission in these channels due to its relatively high brightness, especially at projected regions around the M82 galaxy and in the northern parts of the western arm of the complex where residual emission are seen. This is partly because the spatial region spanned by the M81 system (masked before modelling the MW emission) represents an important part of the area covered by the mosaic. This therefore lowers the accuracy of the constructed model, resulting in a poor subtraction of the MW in certain regions. However, given the limited number of affected channels and regions, this does not significantly impact the overall global \hi\ properties of the system.

\begin{figure*}
\makebox[\textwidth][c]{
\includegraphics[width=0.9\textwidth]{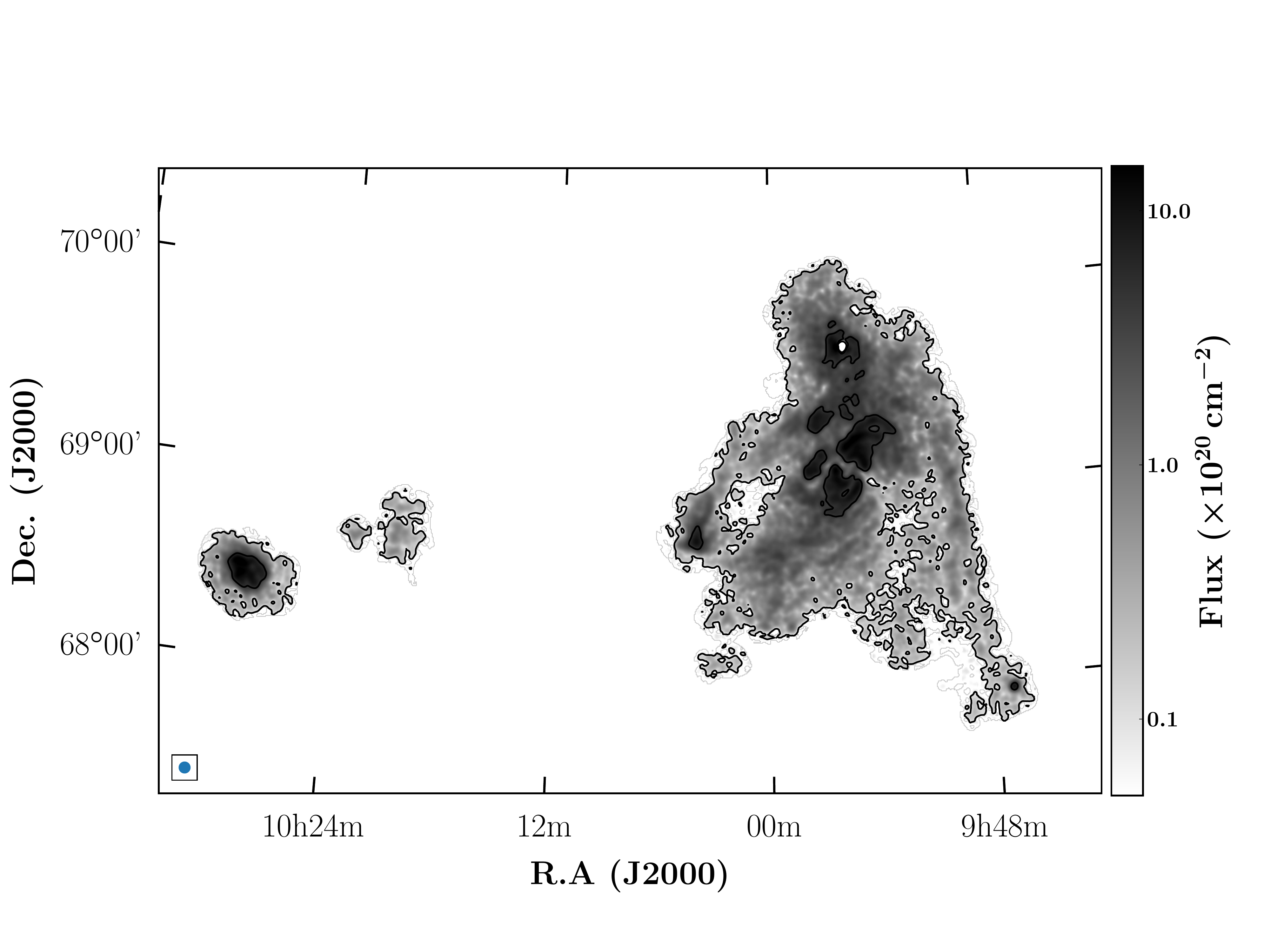}}
\vspace{-20pt}
\caption{Column density map of the data computed by blanking MW channels. The contour shows the $10^{19}\,\cm$ column density level.}\label{fig:nomw}
\end{figure*}

%%%%%%%%%%%%%%%%%%%%%%%%%%%%%%%%%%%%%%%%%%%%%%%%%%%%%%%%%%%%%%%%%%%%%

%%%%%%%%%%%%%%%%%%%%%%%%%%%%%%%%%%%%%%%%%%%%%%%%%%%%%%%%%%%%%%%%%%%%%
%%%%%%%%%%%%%%%%%%%%%%%%%%%%%%%%%%%%%%%%%%%%%%%%%%%%%%%%%%%%%%%%%%%%%
%%%%%%%%%%%%%%%%%%%%%%%%%% HI PROPERTIES %%%%%%%%%%%%%%%%%%%%%%%%%%%%

\section{\hi\, distribution in the system}\label{sec:hi-properties}

\subsection{Major galaxies of the group}
The major three galaxies of the group have been the target of numerous studies in several bands, which helped to shed light on their structure and properties. The general picture of the group is the MW-like galaxy M81 with a stellar mass of $6.4\e{10}\,\Mo$ interacting with M82 and NGC 3077 of stellar masses $1.2\e{10}\,\Mo$ and $2.1\e{9}\,\Mo$ respectively. Their optical properties, as well as those of the other two massive members of the group (NGC 2976 and IC 2574), are listed in \Cref{tb:opticalprops}. Being the most massive galaxy in the group, M81 (NGC 3031) is a prototypical grand design spiral galaxy. Optical HST observations of the galaxy \citep{Chandar2001} have revealed the presence of a young, blue cluster population, whose formation appeared to have begun about 600 Myr ago as a result of the galaxy's interaction with its counterparts. Later, \citet{Davidge2009} found that the specific SFR in the M81 disk is consistent with that expected for an isolated galaxy of a similar mass, indicating that the SFR in M81 has not significantly increased in the past few tens of millions of years.
The second largest galaxy M82 (NGC 3034) is a starburst galaxy at a distance of 3.5 Mpc \citep{Jacobs2009}, host of several reported supernovae \citep{Mattila2013,Gendre2013} including the recently discovered bright, Type Ia supernova SN 2014J \citep{Goobar2014}. It is a well studied edge-on galaxy, classified as an irregular galaxy in the RC3 catalogue \citep{DeVaucouleurs1991} but probably more an SBc \citep{Mayya2005}, and provides a good laboratory to study in great detail the properties of the neutral and molecular gas \citep[e.g.,][]{Yun1993,Neininger1998,Matsushita2005}. \citet{Walter2002} detected molecular streamers in and below the M82's disk, whose structures provide evidence that the molecular gas within the optical disk of the galaxy is disrupted by the interaction with M81. This further confirms an earlier claim by \citet{Sofue1998}, who described M82 as a low-mass galaxy with a strong central concentration of mass and hardly any evidence for a dark matter halo.
The smallest of the three galaxies, NGC 3077, was found to have lost its gas during a close encounter with M81 \citep[e.g.,][]{Brouillet1991,Thomasson1993,Yun1993} and high resolution observations show that as much as 90\% of the galaxy's atomic gas is situated in a prominent tidal arm \citep{Walter2002a}. 

\begin{table*}
	\centering
	\begin{tabular}{l c c c c c c c c c}
	\hline
    \hline
	 Object & \multicolumn{2}{c}{J2000 position} & Type & Distance & $m_{\rm B}$ & $\rm M_\star$ & $d_{\rm maj}\times d_{\rm min}$ & P.A & Incl.\\
	 & R.A & Dec. & & (Mpc) & (mag) & ($10^9\,\Mo$) & ($'$) & ($\dg$) & ($\dg$)\\
	(1) & \multicolumn{2}{c}{(2)} & (3) & (4) & (5) & (6) & (7) & (8) & (9)\\
	\hline
	M81      & 09 55 33.2 & 69 03 55.1 & SAab & 3.63 & 7.89 & 63.8 & $26.9\times14.0$ & $157$ & $57$\\
	M82      & 09 55 52.7 & 69 40 45.8 & SBc$^a$ & 3.53 & 9.30 & 12.1 & $11.2\times4.3$ & $65$ & $82$\\
	NGC 3077 & 10 03 19.1 & 68 44 02.1 & Sm$^b$ & 3.82 & 10.61 & 2.1 & $5.4\times4.5$ & $45$ & $38$\\
	NGC 2976 & 09 47 15.5 & 67 54 59.0 & SAc pec & 3.56 & 10.82 & 1.2 & $5.9\times2.7$ & $143$ & $61$\\
	IC2574   & 10 28 23.5 & 68 24 43.7 & SABm & 4.02 & 10.80 & 0.4 & $13.2\times5.4$ & $50$ & $60$\\
	\hline
	\end{tabular}
	\caption{Optical properties of the main galaxies in the M81 group. Column (1): Galaxy name; Column (2): Optical position from NED; Column (3): Morphological type from RC3; Column (4): Distance to the galaxy from \citet{Karachentsev2004}; Column (5): Total B-magnitude from RC3; Column (6): Stellar mass of the galaxy derived from {\it WISE} photometry (courtesy of Thomas H. Jarrett) at the distance listed in column (4); Column (7): Optical diameter measured at the 25th magnitude from \citet{Karachentsev2004}; Column (8): Optical position angle from \citet{Nilson1973}. Column (9): Optical inclination from \citet{Appleton1981}. Notes: $^a$ morphological type from \citet{Mayya2005}; $^b$ morphological type from \citet{Karachentseva1979}}\label{tb:opticalprops}
\end{table*}

\subsection{Global \hi\, profiles and moment maps}\label{sec:profs_moms}
To compute the global profile of the total \hi\ in the field and the moment maps, we first built a mask of the datacubes. This was done using the smooth and clip algorithm of the {\sc SoFiA} pipeline \citep{Serra2015}. The pipeline takes as input the (MW-subtracted) datacube and a weight map of the field, computed by taking the inverse of the rms noise at each spatial position of the cube. The data presented a long horizontal artefact below the system and east of NGC 2976, so this region was flagged in the masking process. The algorithm creates different convolved versions of the cube, and for each of them builds a mask using a user-defined threshold. These masks are later combined into a single, final mask which, in principle, contains both the bright and low-column density structures. Because of the non-uniformity in the distribution of the noise throughout the cube, we used the median absolute deviation (MAD) method on the negative side of the flux histogram to derive the noise in the data. This method computes the noise from the negative voxels only, which are less likely to contain real emission. We performed extensive tests and found that the MAD method provides a more robust measure of the noise than the standard deviation. We also found that the optimum mask is obtained with a combination of six smoothing kernels, which convolve each the data to resolutions equal to 1, 2 or 3 times the original resolution along either the spatial or spectral axes. The convolution process of the data uses a Gaussian kernel along the spatial axes, and a boxcar kernel along the spectral axis. The detection threshold at each of these resolutions is $7\sigma$, where $\sigma$ is the noise level at the given resolution determined using the above-mentioned MAD method. During the merging phase of the detected pixels into individual sources, we have required that only sources spanning at least 10 channels (33 \kms) be retained. This ensures that noise peaks spanning a few channels are discarded. An additional test was made with a lower velocity constraint of 23 \kms, but no new feature was discovered. Furthermore, we have used the reliability module of the package \citep[illustrated in][]{Serra2012a} to evaluate the reliability of the different sources from the distribution of their ``positive'' and ``negative'' detections. The positive (negative) detections are defined as having a total positive (negative) total flux. Then we required that sources with a reliability less than 90\% be discarded. %The resulting mask, which only contains reliable reliable sources, was used to compute the global integrated profiles and the moment maps.
A masked version of the datacube was created by applying the resulting mask to the data, and the total \hi\ intensity map (zeroth moment), the intensity-weighted velocity field (first moment) and the velocity dispersion map (second moment) were computed from the masked data using the \moment\ task in \miriad\, \citep{Sault1995}. Also, the global \hi\ profiles of the field were computed by integrating all fluxes within each plane of the masked data.

\begin{figure*}
\centering
\includegraphics[width=0.6\columnwidth]{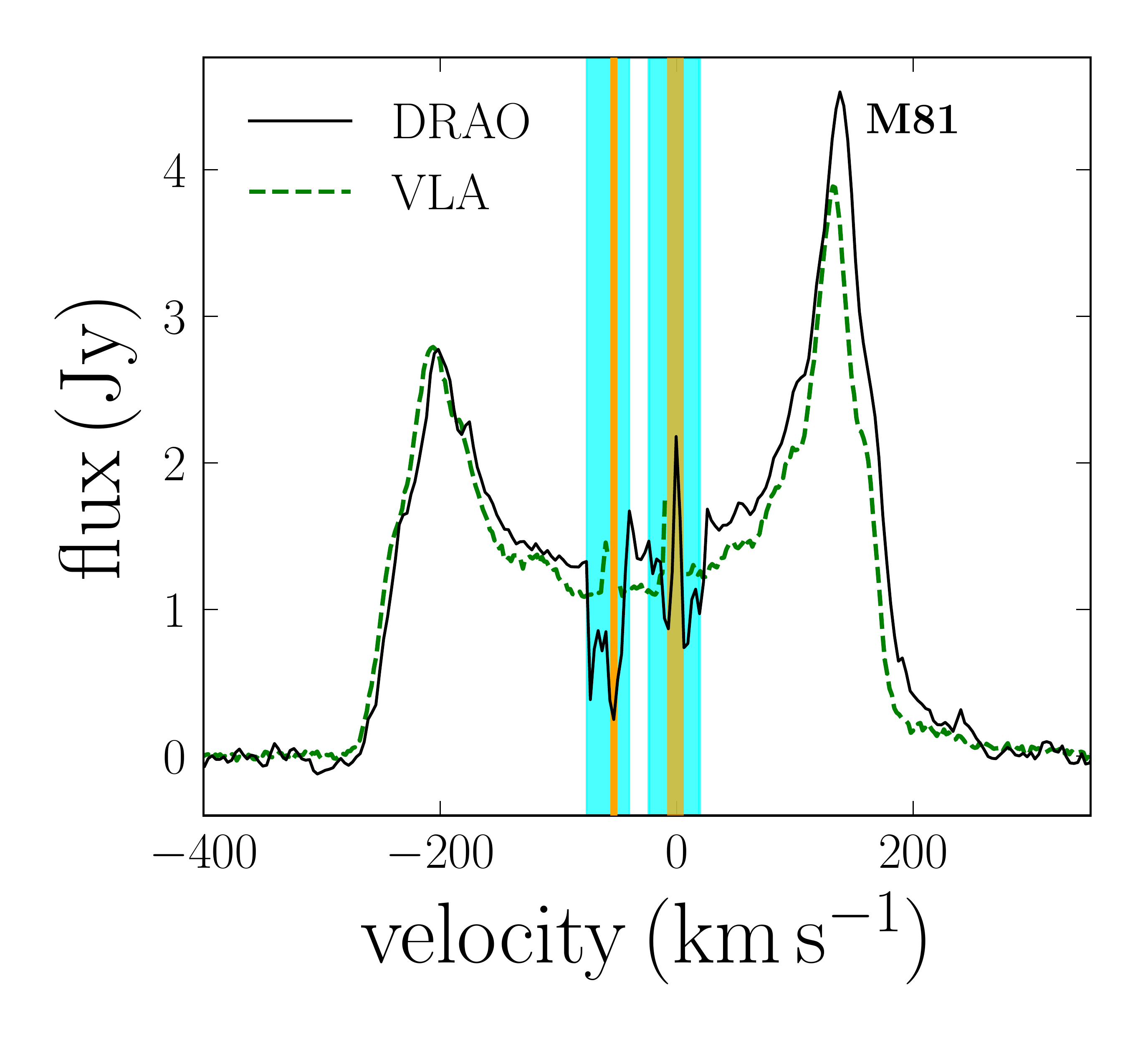}
\includegraphics[width=0.6\columnwidth]{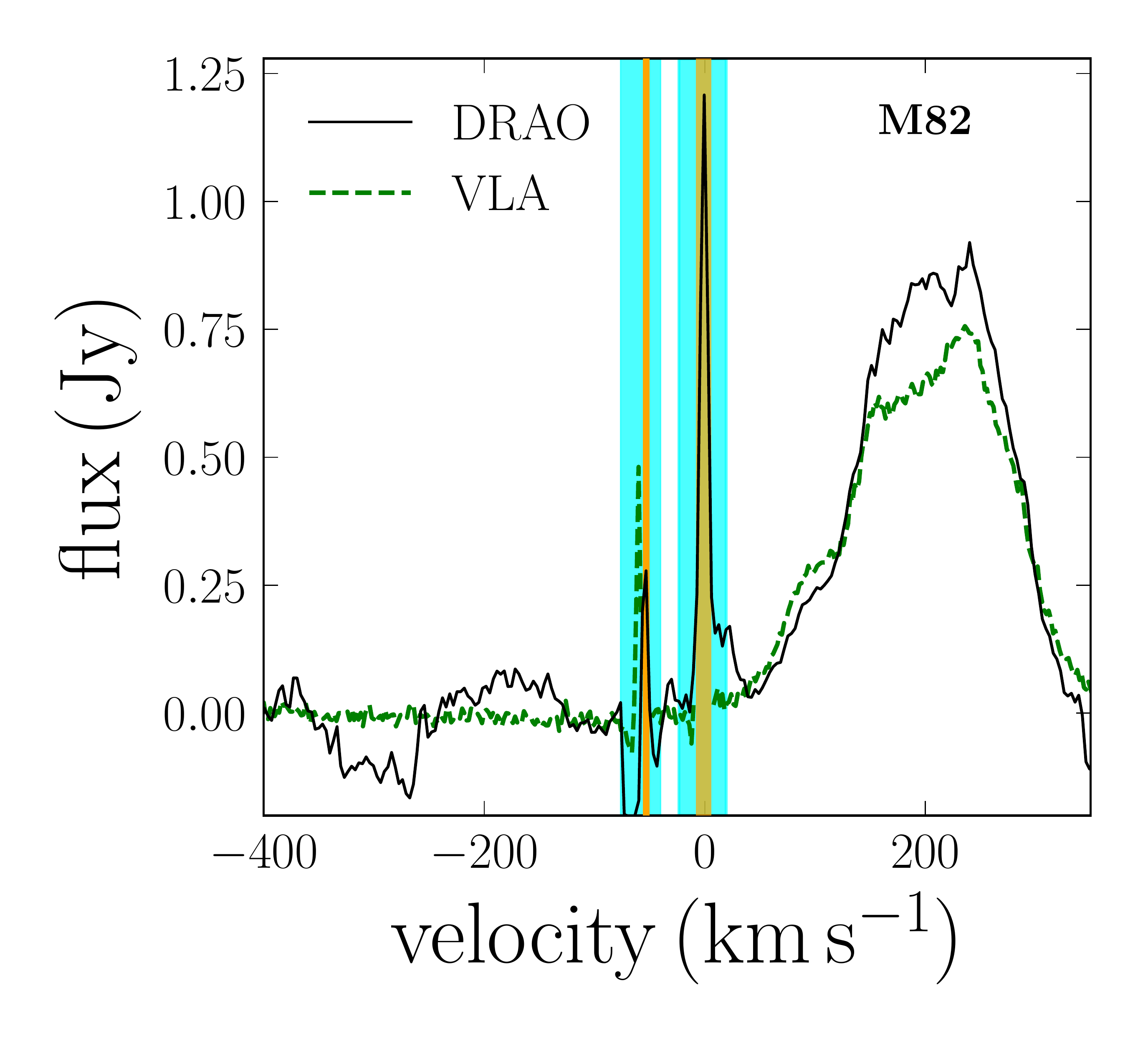}
\includegraphics[width=0.6\columnwidth]{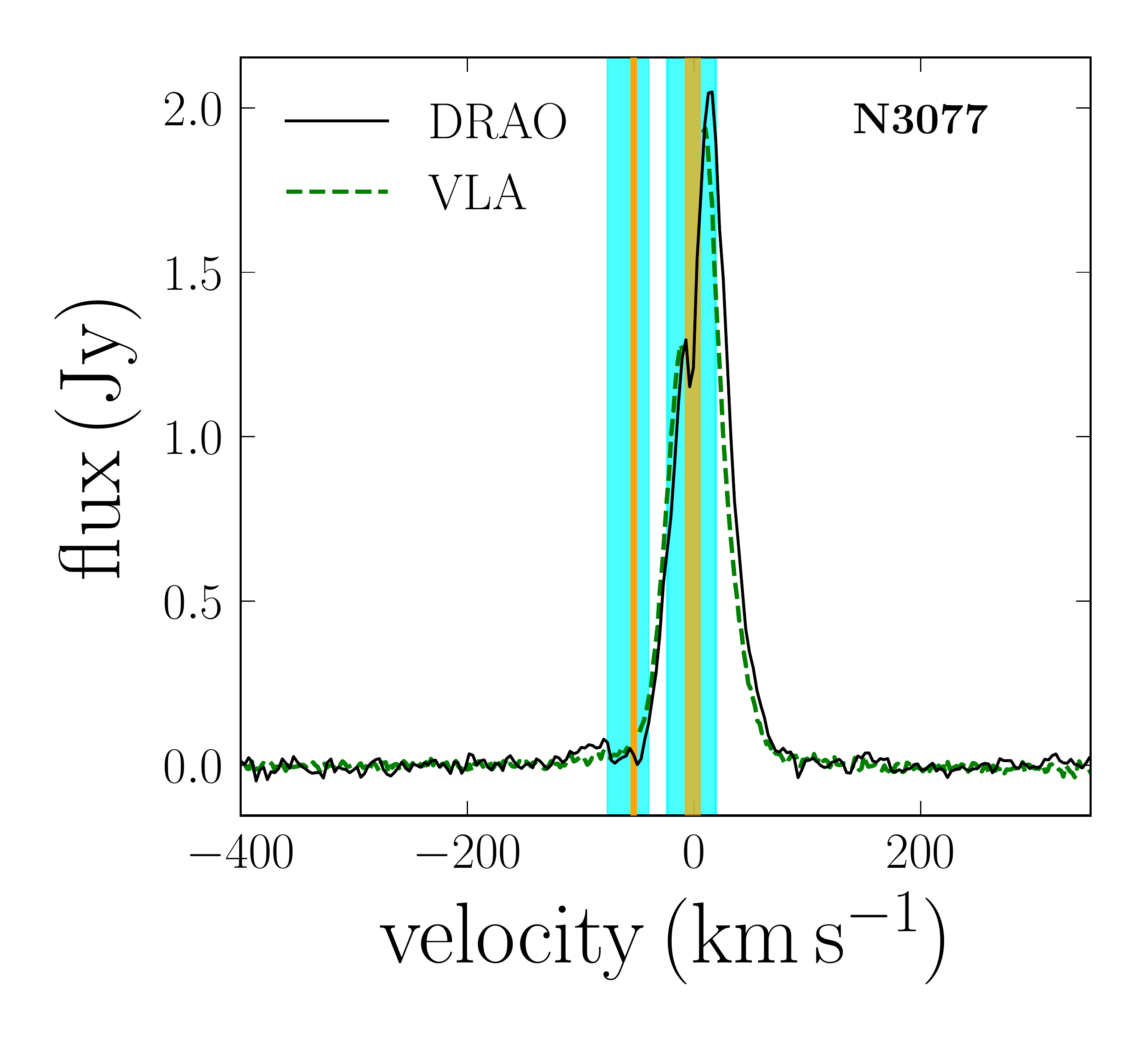}
\includegraphics[width=0.6\columnwidth]{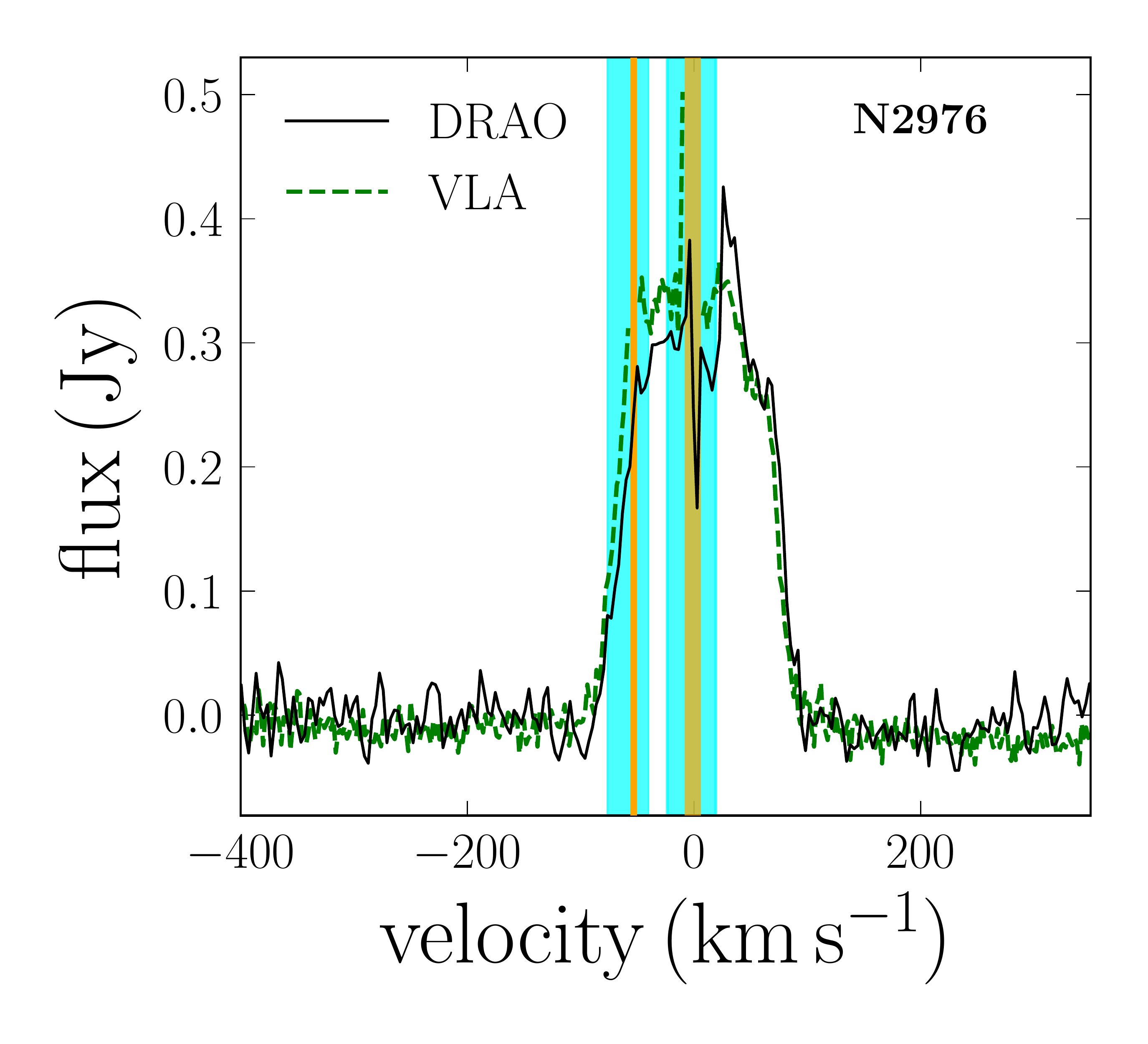}
\includegraphics[width=0.6\columnwidth]{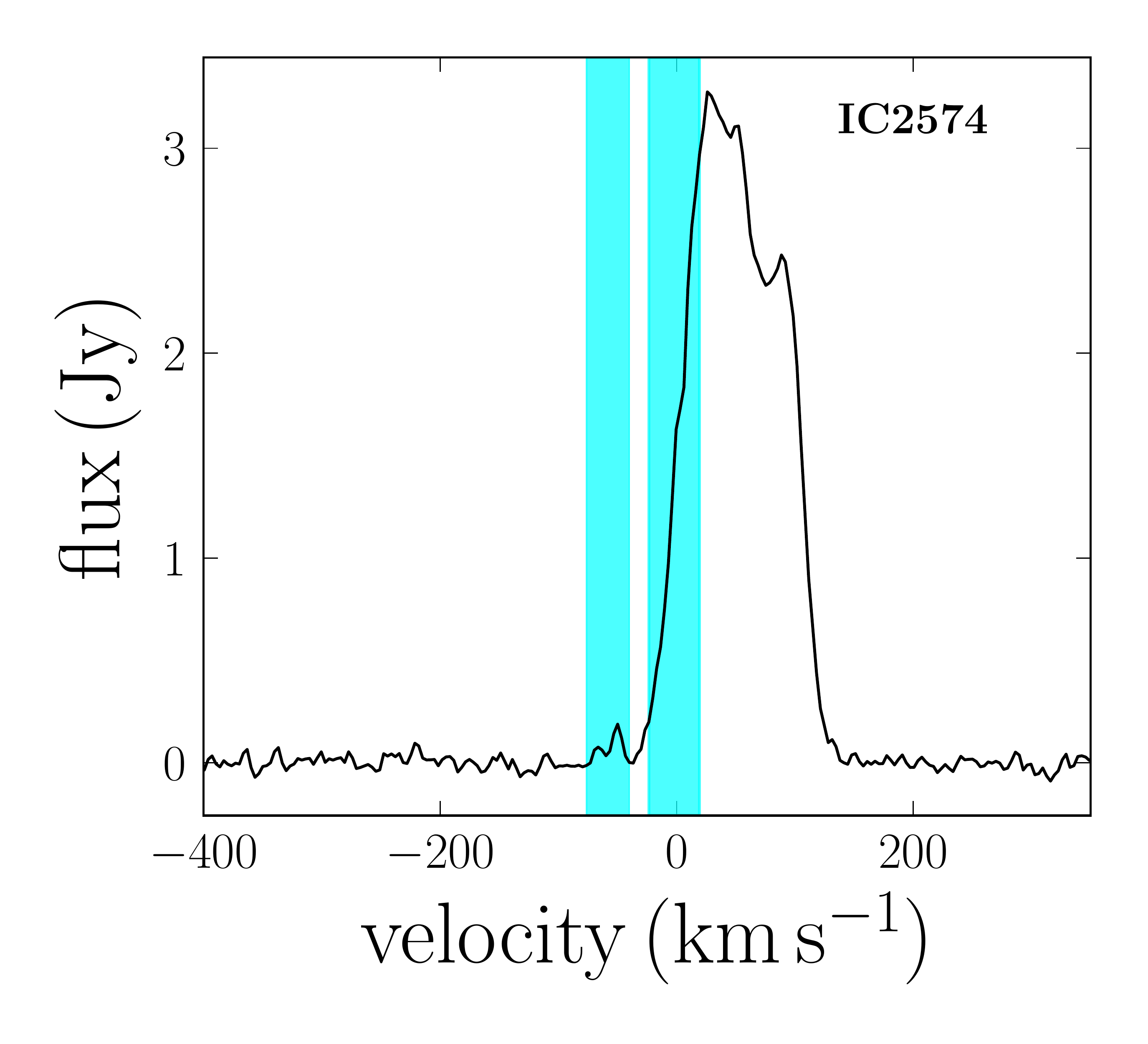}
\includegraphics[width=0.6\columnwidth]{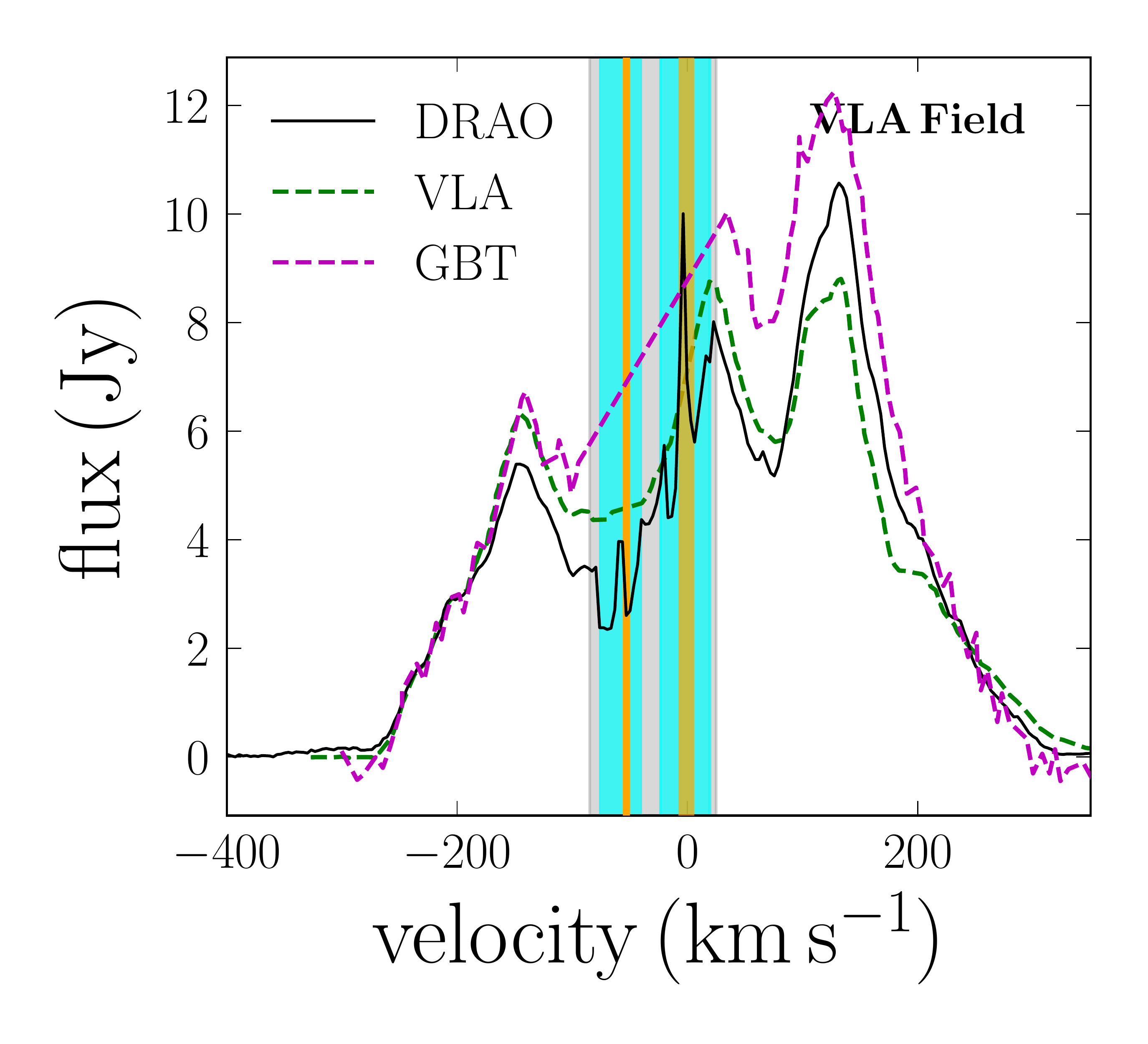}
\includegraphics[width=0.6\columnwidth]{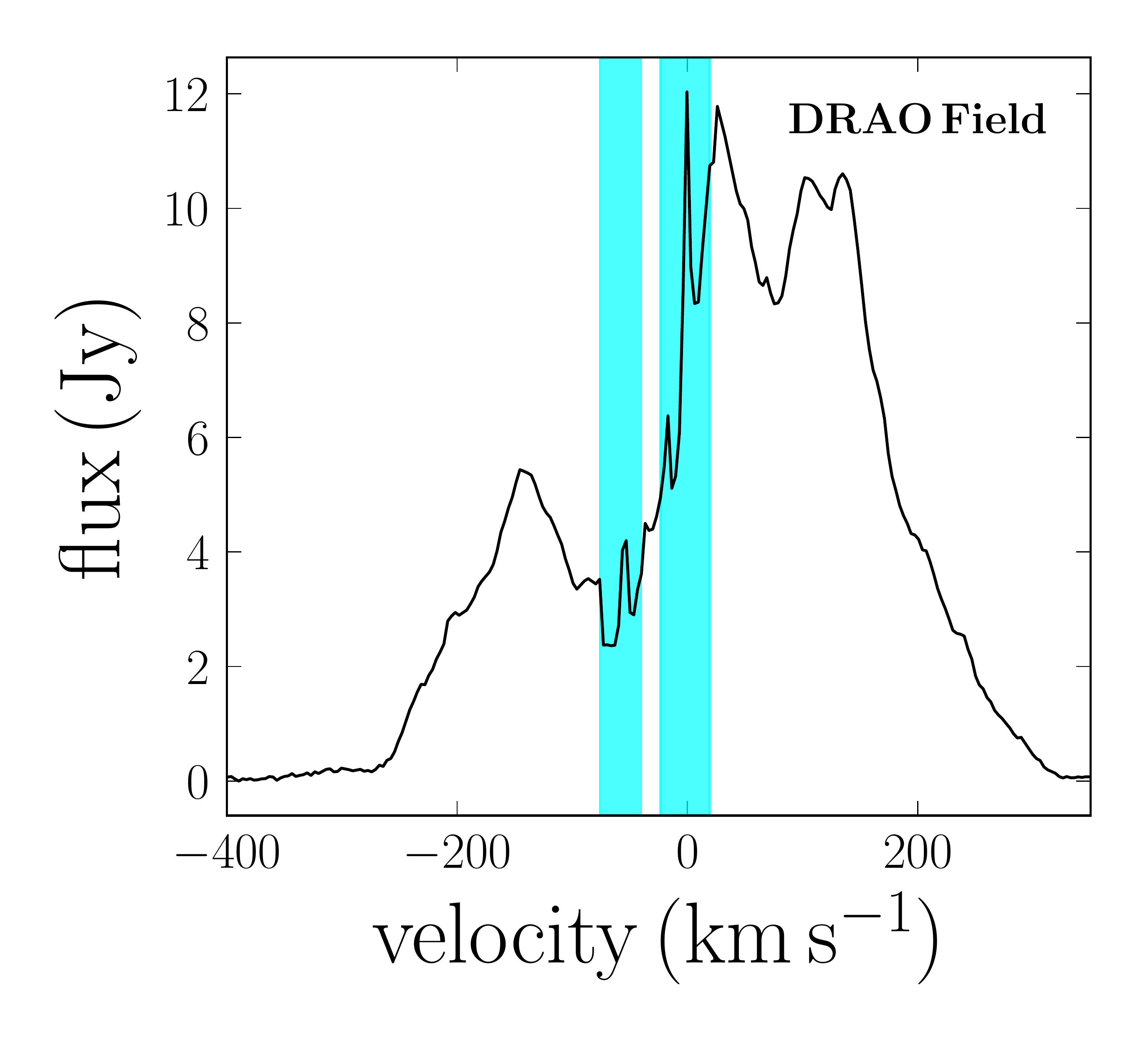}
\caption{\hi\ intensity profiles of the major galaxies of the M81 group (first five panels), derived from the DRAO and VLA D-array datacubes (when available). The last two panels show respectively the \hi\ profiles of the area covered by the GBT (and VLA) observations, and the profile of the entire DRAO field. We also overplot on the second last panel the GBT profile. The vertical shaded areas show the velocity range of the DRAO ({\it cyan shade}), the VLA ({\it orange shade}) and the GBT ({\it light gray shade}) data contaminated by galactic foreground.}\label{fig:hiprofiles}
\end{figure*}

In \Cref{fig:hiprofiles} we show the integrated \hi\ profiles of each of the major galaxies in the field-of-view; for comparison, we overplot in the figure the \hi\ profiles (of the galaxies in the VLA field-of-view) derived from the VLA D-array data. The profiles were extracted within ellipses corresponding to the optical size of the galaxies. We also show in the figure, a comparison of the \hi\ profiles of the DRAO mosaic, VLA D-array mosaic \citep{DeBlok2018a} and GBT \citep{Chynoweth2008} observations, integrated over the same area: the field-of-view of the VLA mosaic. The profiles show an agreement between the DRAO and VLA data for most of the galaxies, except for M82 where DRAO recovers more flux than the VLA. This is because, unlike the VLA D-array data, the DRAO data is zero-spacing corrected and is therefore more sensitive to large-scale, faint structures that exist around M82. It is worth noting that the VLA+GBT mosaic of \citet{DeBlok2018a} also showed the presence of diffuse gas around the galaxy.
Because of the continuum at the centre of M82, any emission or absorption at the core of the galaxy is not reliable; we therefore blanked the inner $2'$ of the galaxy prior to deriving the \hi\ profile. In fact, the bright core of M82 seemed to have phase errors associated with it, which causes baseline variations in the data. Since this is not well modelled by the linear continuum baseline that was fitted and subtracted from the data, the values of the fluxes in this part of the galaxy were found to be inaccurate.

\begin{figure*}
\makebox[\textwidth][c]{
\hspace{30pt}
\includegraphics[width=1.15\textwidth]{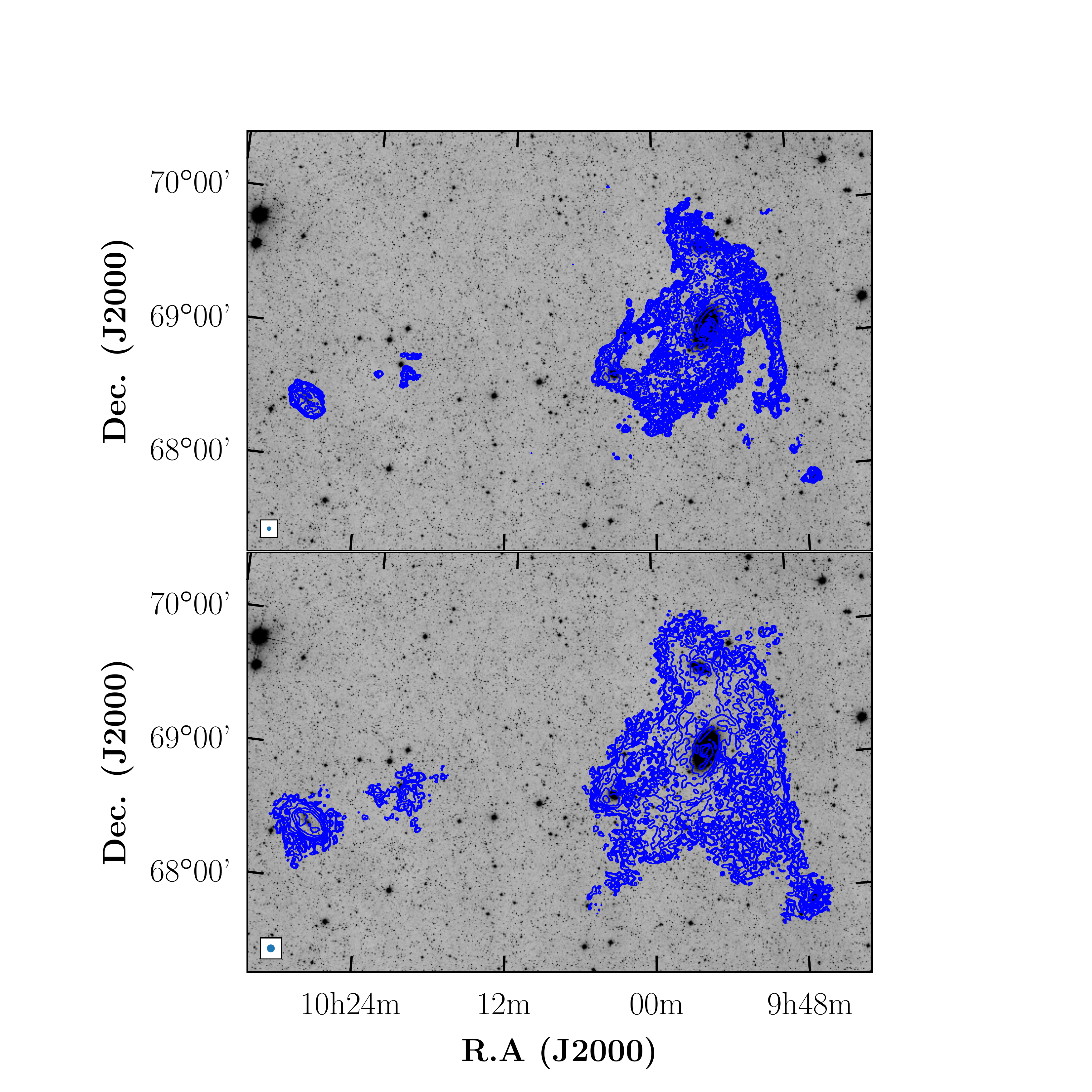}}
\vspace{-20pt}
\caption{Column density maps of the full resolution ({\it top}) and smoothed $1.8'$ ({\it bottom}) DRAO data overlaid on optical {\it WISE} W1 grayscale image. Contours are $1, 2, 4, \dots, 128\e{19}\,\cm$}\label{fig:nhimap_drao}
\end{figure*}

The total \hi\ intensity map of the system, computed from the full resolution and smoothed $1.8'$ resolution datacubes, are presented in \Cref{fig:nhimap_drao}. The western arm of the system is found to extend as far as to NGC 2976. The projected length of the arm is $\sim83'$, which corresponds to a physical size of $\sim87.7$ kpc at the distance of M81, and has an \hi\ mass of $(4.9\pm0.3)\e{8}\,\Mo$.
Moreover, numerous intergalactic \hi\ clumps were detected in the region between the ``main body'' of the system and the arm. The previous GBT map revealed an extended cloud of gas in the region (the so-called ``Cloud 3'' of \citealt{Chynoweth2008}), and the subsequent VLA D-array map of \citet[][Fig. 11 therein]{DeBlok2018a} also showed \hi\ clouds at that position, connecting to the arm. What is new in the present map is that more small clouds are detected in the area, and they tend to fill the space between the arm and the south-western side of M81.

\begin{figure*}
\centering
\makebox[\textwidth][c]{
\includegraphics[width=1.1\textwidth]{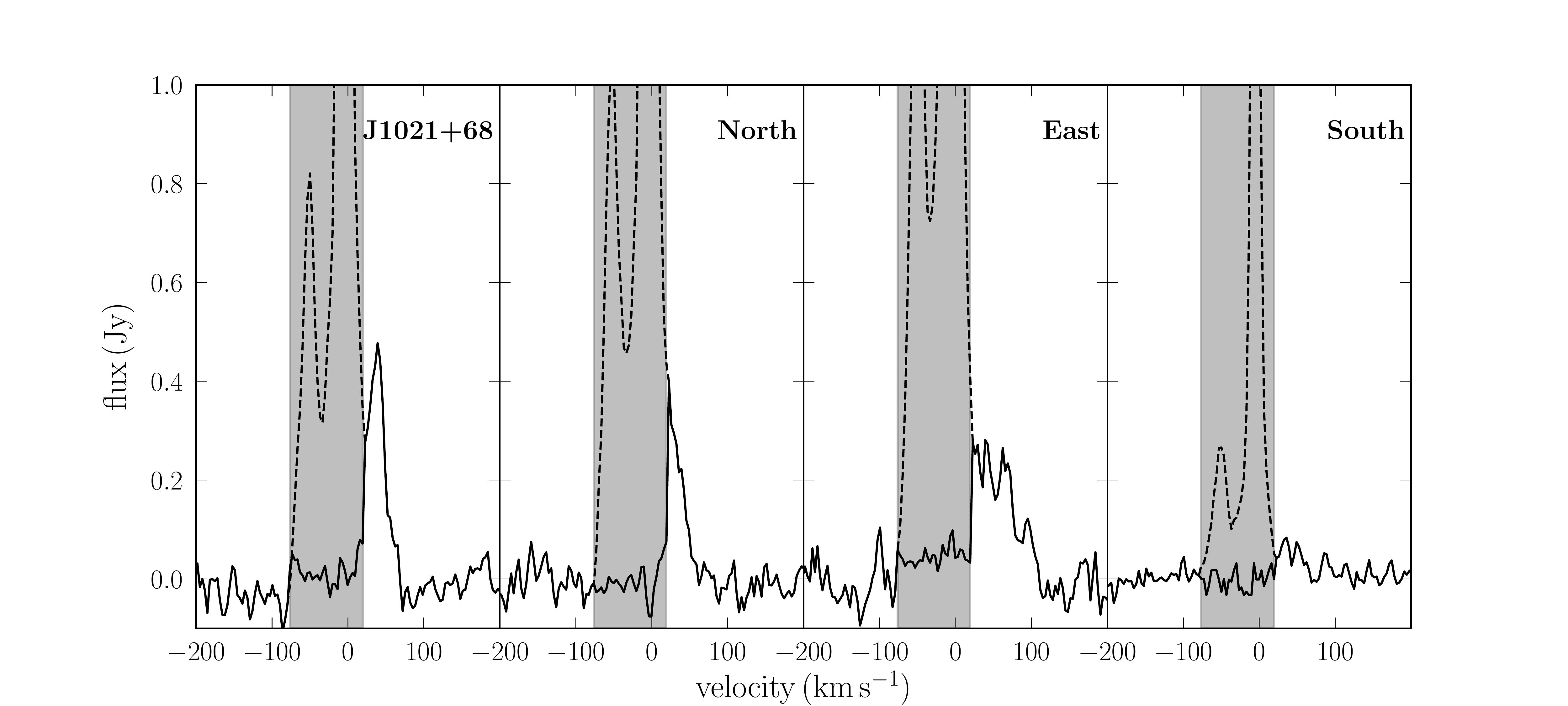}}
\vspace{-10pt}
\caption{The \hi\ profile of HIJASS J1021+68 (first panel) and those of the three neighbouring clouds (second to fourth panel) derived from the MW-subtracted datacube. The {\it dashed line} shows the global profile before MW subtraction. The grey shaded area represents the entire velocity range processed for MW subtraction.}\label{fig:hijass-clouds}
\end{figure*}

\begin{figure*}
\centering
\makebox[\textwidth][c]{
\includegraphics[width=1.1\textwidth]{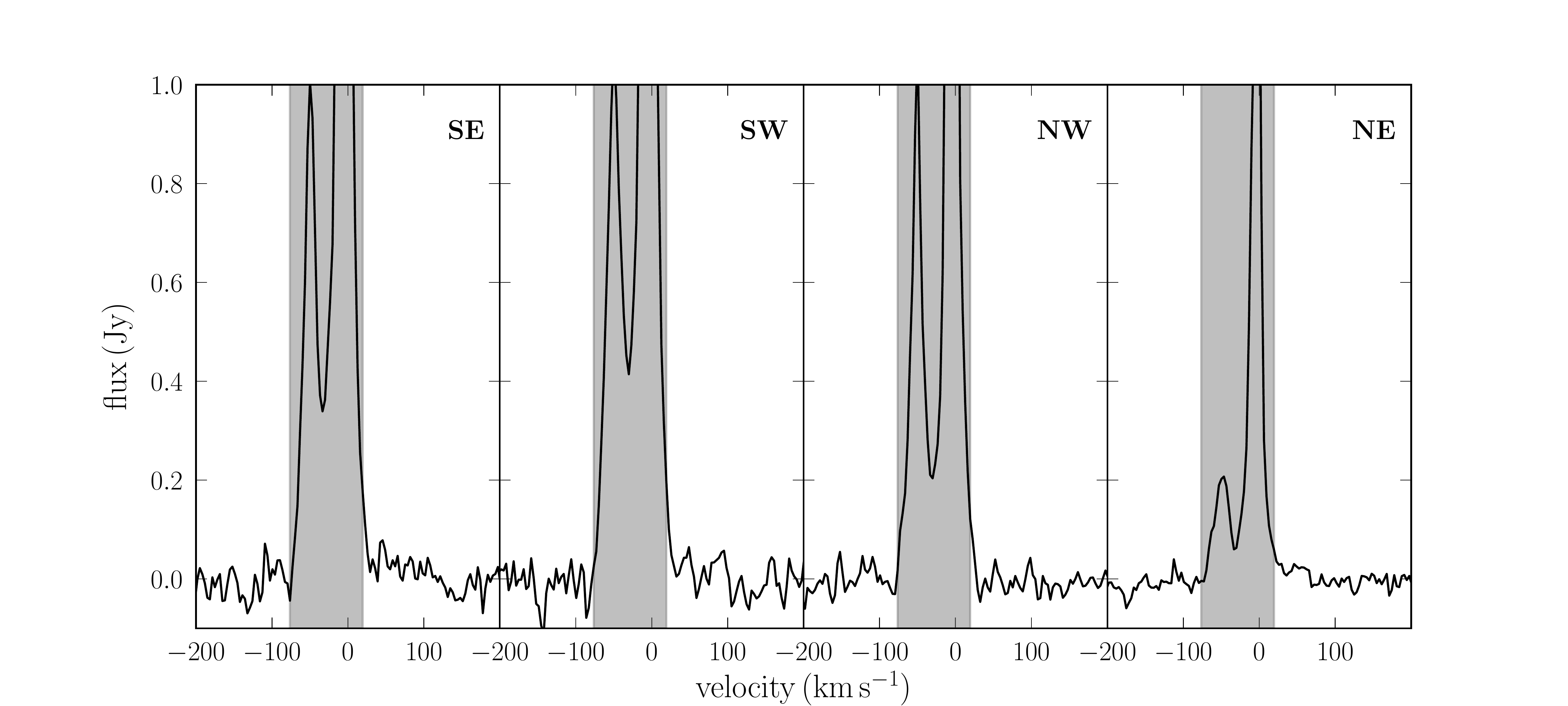}}
\vspace{-10pt}
\caption{The ``control'' \hi\ profiles of boxes taken southeast (SE), southwest (SW), northwest (NW) and northeast (NE) around HIJASS J1021+68 (see \Cref{fig:hijass-boxes} for the positions of the boxes). The peaks of the MW emission in the panels are lower than in the first three panels of \Cref{fig:hijass-clouds} due the relatively smaller size of the ``control boxes''. None of the panels present a profile that spreads outside the grey area, as it the case for the profiles of the clouds in \Cref{fig:hijass-clouds}.}\label{fig:control-specs}
\end{figure*}

\begin{figure}
\centering
\makebox[\columnwidth][c]{
\hspace{-30pt}
\includegraphics[width=1.1\columnwidth]{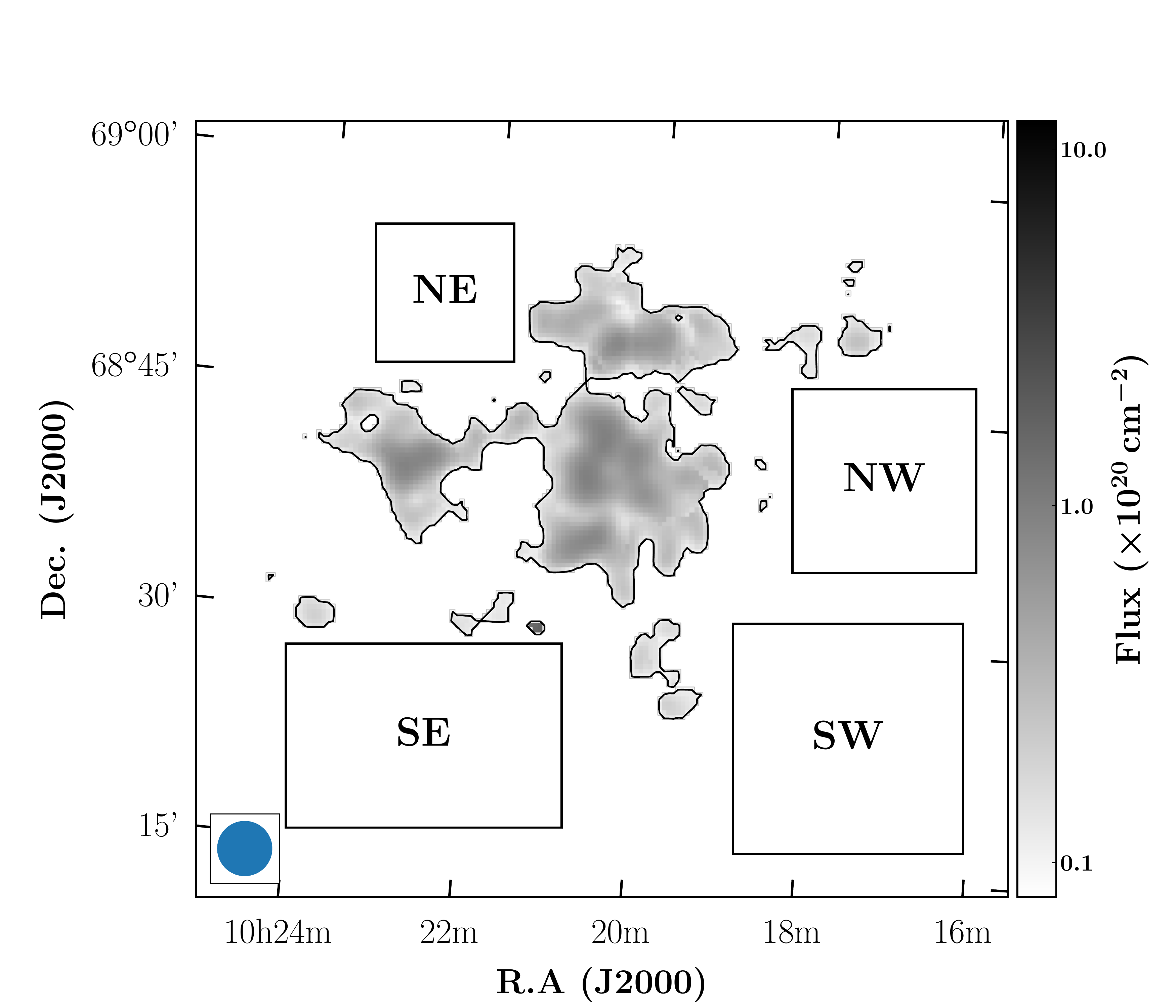}}
\vspace{-10pt}
\caption{A zoom-in on the HIJASS J1021+68 region showing the boxes used to extract the ``control'' profiles of \Cref{fig:control-specs}. The labels of the boxes are the same as in \Cref{fig:control-specs}, and the contour shows the $10^{19}\,\cm$ column density level.}\label{fig:hijass-boxes}
\end{figure}

The primary aim of the present DRAO observations is to make a complete census of \hi\ clouds around the M81 system, and in the space between the system and IC 2574. At a projected distance of 46.5 kpc west of IC 2574, we detected the previously catalogued dwarf galaxy HIJASS J1021+68 \citep{Boyce2001,Lang2003} which is not optically detected. The galaxy, which was previously observed to be a single concentration, appears to be surrounded by \hi\ clouds of various sizes. It also appears to be connected to IC 2574, as was suggested by \citet{Boyce2001}. However, unlike the single-dish observations, the high resolution DRAO mosaic allows to resolve the connection between the two galaxies, and it is clear in \Cref{fig:nhimap_drao} that there exists a ``filament''-like complex of clouds connecting the two galaxies. In \Cref{fig:hijass-clouds} we show the \hi\ profile of HIJASS J1021+68, as well as those of the neighbouring clouds. Although the systemic velocities of the clouds -- and that of the dark galaxy as well -- coincide with the MW emission, the profiles show that they are located in spectral regions that are not greatly affected by MW emission and therefore did not require a MW subtraction.
However, the position of the peak emission of these clouds with respect of the MW emission suggests that they may just be wings of broader MW emission. To ensure that these structures are not just an extension of the Galactic emission, we have extracted the profiles of four ``control'' regions of various sizes, at various positions around HIJASS J1021+68. These ``control'' profiles are shown in \Cref{fig:control-specs}, and the respective regions they are extracted from are represented by boxes in \Cref{fig:hijass-boxes}. Contrarily to the profiles of HIJASS J1021+68 and associated clouds, the control spectra show no presence of broad \hi\ emission extending outside the MW-subtracted channels, showing that the clouds are most likely real and are not just remnant Galactic emission.

The derived \hi\ mass of HIJASS J1021+68 is $(3.5\pm0.7)\e{7}\,\Mo$, in agreement with the quoted HIJASS mass of $3\e{7}\,\Mo$. For the complex of small \hi\ clouds lying east of the object, which seem to form a ``filament'' towards IC 2574, we measured a total \hi\ mass of $(1.3\pm0.5)\e{7}\,\Mo$. The northern cloud of HIJASS J1021+68, which seems to connect to the dark galaxy, has an \hi\ mass of $(1.6\pm0.6)\e{7}\,\Mo$, and the cloud south to the galaxy has a mass of $(0.1\pm0.2)\e{7}\,\Mo$.
A more detailed study of this eastern region of the M81 group will be presented in a subsequent paper. 

\begin{figure*}
\centering
\includegraphics[width=0.9\textwidth]{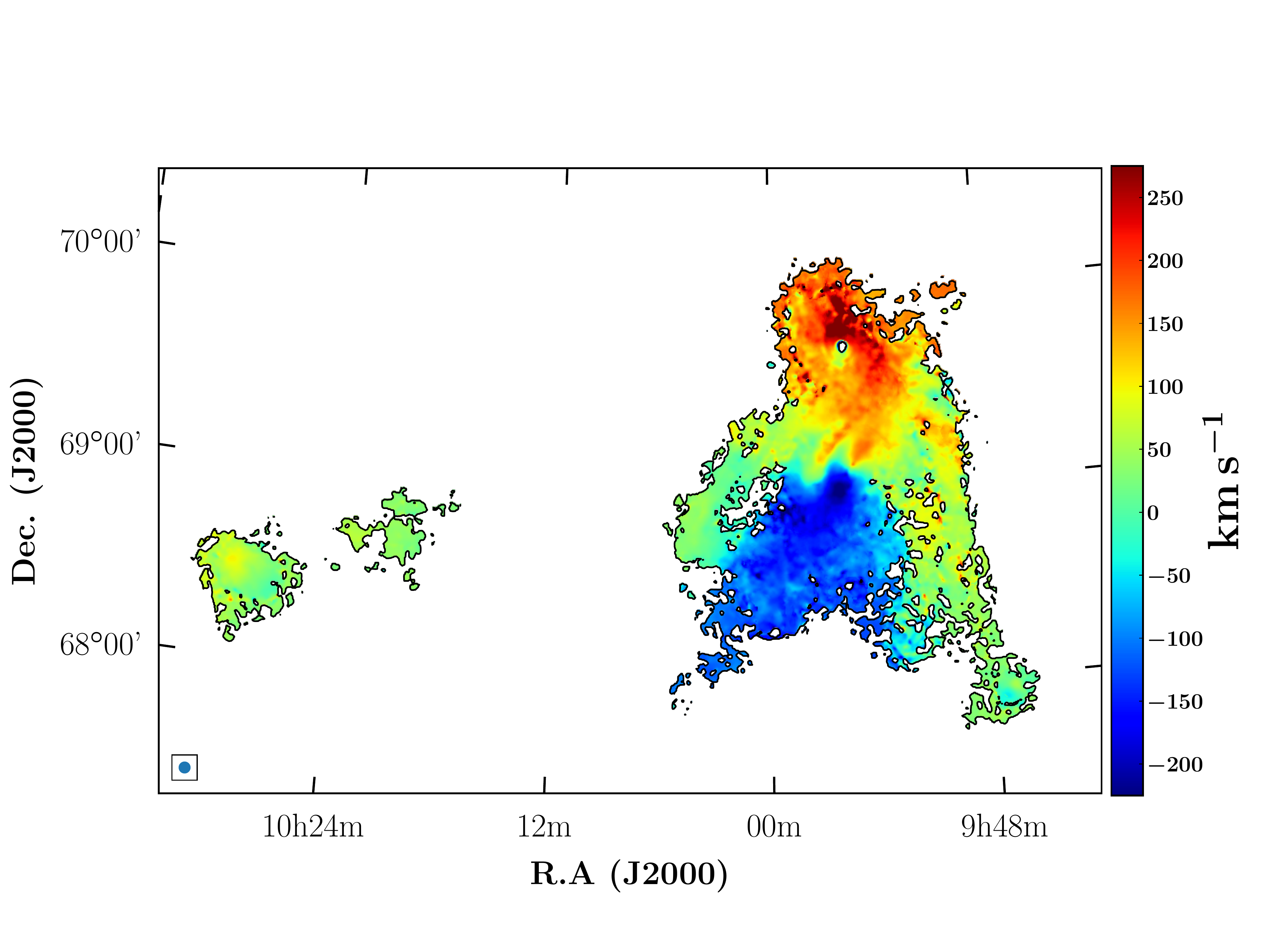}
\vspace{-20pt}
\caption{The intensity weighted velocity field of the M81 group derived from the DRAO $1.8'$ resolution datacube. The contour corresponds to the $1\e{19}\,\cm$ column density level.}\label{fig:vfield_drao}
\end{figure*}

In \Cref{fig:vfield_drao} we present the velocity field of the complex, computed from the DRAO datacube with the \miriad\, task \moment. As noted in \citet{DeBlok2018a}, the velocity field shows a regular rotation of the disk, especially in the inner regions. Beyond a certain radius (approximately 10 kpc along the minor axis) the rotation of the galaxy is disrupted, especially along its minor axis and towards the position of the dwarf galaxy HoIX. There is little velocity gradient along the western arm, implying that the arm is almost contained in the plane of sky. We also note an arc connecting HoIX to M82, hinting at a kinematical association between the two galaxies. There seems to be a velocity continuity between the western side of IC 2574, the gas cloud HIJASS J1021+68, and the region of the system containing NGC 3077 and the eastern bridge of the system. This, together with the fact that there exists a peak of an apparent cloud in the space between HIJASS J1021+68 and the NGC 3077, may suggest that IC 2574 is not completely isolated from the system.

\begin{figure*}
\centering
\includegraphics[width=0.9\textwidth]{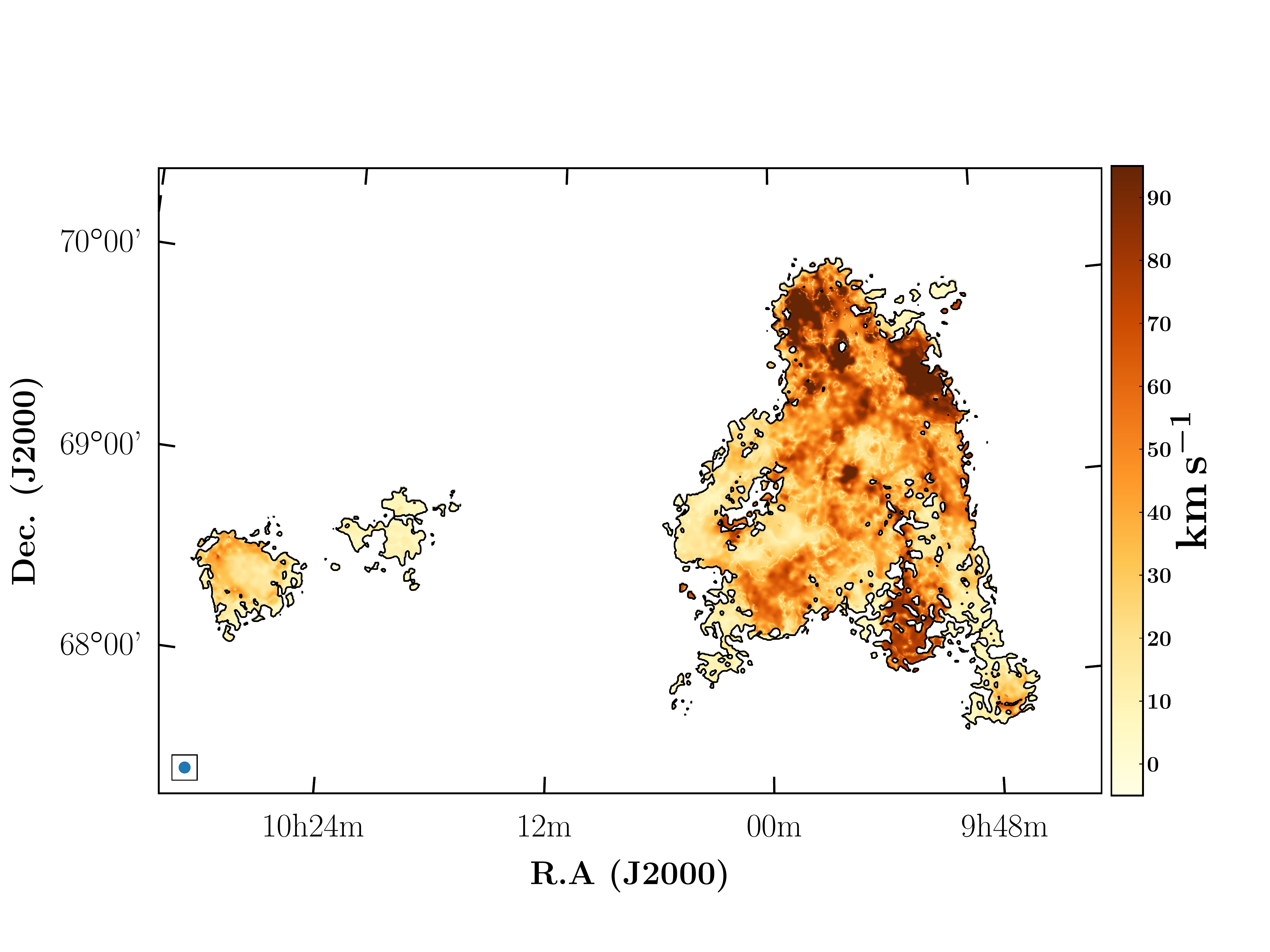}
\vspace{-20pt}
\caption{The velocity dispersion map of the M81 group derived from the DRAO $1.8'$ resolution datacube. The contour corresponds to the $1\e{19}\,\cm$ column density level.}\label{fig:vdisp_drao}
\end{figure*}

The velocity dispersion map of the group, presented in \Cref{fig:vdisp_drao}, shows that the velocity dispersion is, as expected, highest around M82 and in the northernmost region of the western arm. While the dispersion may be intrinsically high around M82, a comparison of the velocity dispersion map with that obtained from the blanked MW channels data has suggested that the high dispersion value in the upper region of the western arm and in the region northeast of M82 is undoubtedly artificial and is likely a result of the MW subtraction.

\subsection{\hi\ masses}
The overall \hi\ detected in the field of the VLA observations, encompassing the M81 system as well as NGC 2976, has a total \hi\, mass of $1.1\e{10}\,\Mo$ (at the distance of M81) as derived from the DRAO data.
As discussed in \citet{DeBlok2018a}, determining the \hi\, mass of the individual galaxies in the complex can only be done with a limited accuracy, due to the presence of tidal \hi\, features that make it impossible to determine their true \hi\, extent. Several authors \citep{Appleton1981,Yun1999,Chynoweth2008,DeBlok2018a} who have determined the \hi\, mass of the complex's members have performed the measurement within the optical disks of the galaxies. Although this method is likely to underestimate the \hi\, masses (given that the \hi\, envelope is known to extend father than the optical disk), it provides, at the moment, the best estimate that one can obtain. We use here the optical size ($D_{25}$) of the different galaxies to derive their \hi\, mass, which we compare in \Cref{tb:himass} to the values in the literature. We find that the masses derived in this work are higher than those of the VLA D-array of \citet{DeBlok2018a}, but lower than those of \citet{Chynoweth2008}. This is consistent with the expectations, given that the DRAO data is corrected for short-spacing unlike the VLA-only data. Also, the values of \citet{Chynoweth2008} are not only derived from a single-dish telescope, but the fluxes of the sources in the MW-affected channels of the GBT \hi\ datacube are a linear interpolation of the fluxes outside those channels.
The masses of \citet{Yun1999} are systematically higher than those measured in the present work, in \citet{DeBlok2018a}, and in \citet{Chynoweth2008}. 
It is not clear what ellipses \citet{Yun1999} measured the \hi\ masses in, and what the method used for the MW \hi\ subtraction is. To get an estimate of the uncertainty in the \hi\ masses, \citet{DeBlok2018a} measured the masses within a radius of $2R_{25}$ and their obtained masses are still lower than those of \citet{Yun1999}. The method adopted to determine the masses of the individual galaxies due to the complexity of the system is highly dependent on the ellipses considered, and the different definitions of the size of the galaxies constitute a source of discrepancy between the different \hi\ masses.

Also, it is not clear what method the authors used to mitigate the foreground galactic \hi\ in their data, which covers a non-negligible part of the M81 group. Besides M81, the \hi\ masses of all galaxies derived in \citet{Appleton1981} are higher than those derived in this work, which may be due to the reasons mentioned above.

The total \hi\ mass of the galaxies in the M81 system is $3.5\e{9}\,\Mo$, whereas the total \hi\ mass detected in the field is $1.1\e{10}\,\Mo$. This implies that only about 31\% of the \hi\ in the region resides in galaxies. The majority of the \hi\ is in the form of intergalactic gas and ``lives'' in structures like bridges, tails and clouds.

\begin{table}
	\centering
	\begin{tabular}{l c c c c c}
	\hline
    \hline
	\multirow{2}{*}{Galaxy} & \multicolumn{5}{c}{$M_\HI$ ($\times10^9\, \Mo$)} \\
	& DRAO & dB18 & C08 & Y99 & A81\\
	\hline
	M81      & 2.54 & 2.29 & 2.67 & 2.81 & 2.17\\
	M82      & 0.44 & 0.44 & 0.75 & 0.80 & 0.72\\
	NGC 3077 & 0.33 & 0.23 & 1.01 & 0.69 & 1.00\\
	NGC 2976 & 0.14 & \dots   & 0.52 & \dots & 0.16\\
	Total    & 11.2 & 7.74 & 10.46 & 5.6 & 5.4\\
	IC2574   & 1.4 & \dots & \dots & \dots & 1.0\\
	Total+IC2574 & 12.6 & \dots & \dots & \dots & 6.4\\
	\hline
	\end{tabular}
	\caption{\hi\, masses of the main galaxies in the M81 group. The DRAO masses were derived at the M81 distance. dB18 = \citet{DeBlok2018a}; C08 = \citet{Chynoweth2008}; Y99 = \citet{Yun1999}; A81 = \citet{Appleton1981}.}\label{tb:himass}
\end{table}

\subsection{Zero-spacing corrected VLA data}
The zero-spacing corrected version of the C+D array data presented in \citet{DeBlok2018a} was obtained using a GBT dataset from \citet{Chynoweth2011}. This GBT dataset covers the M81/M82 and NGC 2403 groups, and has a sensitivity of $2.5\e{17}\,\cm$ (at a spatial resolution of $9.1'$) in the area covering the system. However, due to the baseline subtraction procedure used for the GBT data, the zero-spacing corrected VLA cube contains many artefacts in the Galactic velocity range. This makes it hard to apply the MW subtraction technique described in \Cref{sec:mwremoval}, and the analysis in \citet{DeBlok2018a} excludes the MW-affected velocity range.
Because of the poor overlap between the two datasets in the {\it uv} plane (35m to 90m), the data also shows negative bowls around emission peaks, and especially around M82.
We therefore re-applied the zero-spacing correction to the VLA data using the Effelsberg EBHIS data of the area covering the M81 galaxies, in order to improve the detection of gas in that region with respect to the analysis made in \citet{DeBlok2018a}. The EBHIS survey is properly Nyquist sampled on the sky, and can be considered ``complete'' in the sense that it observes all spatial structures down to its resolution limit. Also, the baseline subtraction procedure was done in a better way than the GBT data, such that there are no negative artefacts in the EBHIS data of the M81 region.

The resulting data contains less artefacts, and shows an improvement over the GBT zero-spacing data. The datacube is $2048\times2048\times450$ pixels in size, with a pixel size of $6''$ and a channel spacing of 2 \kms.
Applying the technique described in \Cref{sec:mwremoval}, we subtracted the MW emission from the datacube, and used once again the smooth and clip algorithm implemented in SoFiA to build a mask of the resulting datacube. For such a large size of data, SoFiA requires an unusual amount of computing power to work efficiently \citep[see][]{Serra2015}. However, the beam size of the data is $38.1''\times30.9''$, translating to a sampling of more than 6 pixels per beam (along the beam's major axis). We therefore spatially resampled the cube to about 3.2 pixels per beam, decreasing its size to $1024\times1024\times450$ pixels. This was done with the \miriad\ task {\sc Regrid}.
The optimal mask, sensitive to low column density emission and extended structures, was obtained from a combination of masks at a $4\sigma$ clipping level at resolutions (1,1,1), (1,1,2), (2,2,1), (2,2,2) and (3,3,1), where the first two numbers of each triplet represent the multiples of the two spatial resolutions, and the third number the multiple of the velocity resolution. In the masking process, we flagged an aliasing effect that was contained in the VLA mosaic, and appeared East of NGC 3077.

\begin{figure*}
\vspace{-130pt}
\makebox[\textwidth][c]{
\hspace{50pt}
\includegraphics[width=1.15\textwidth]{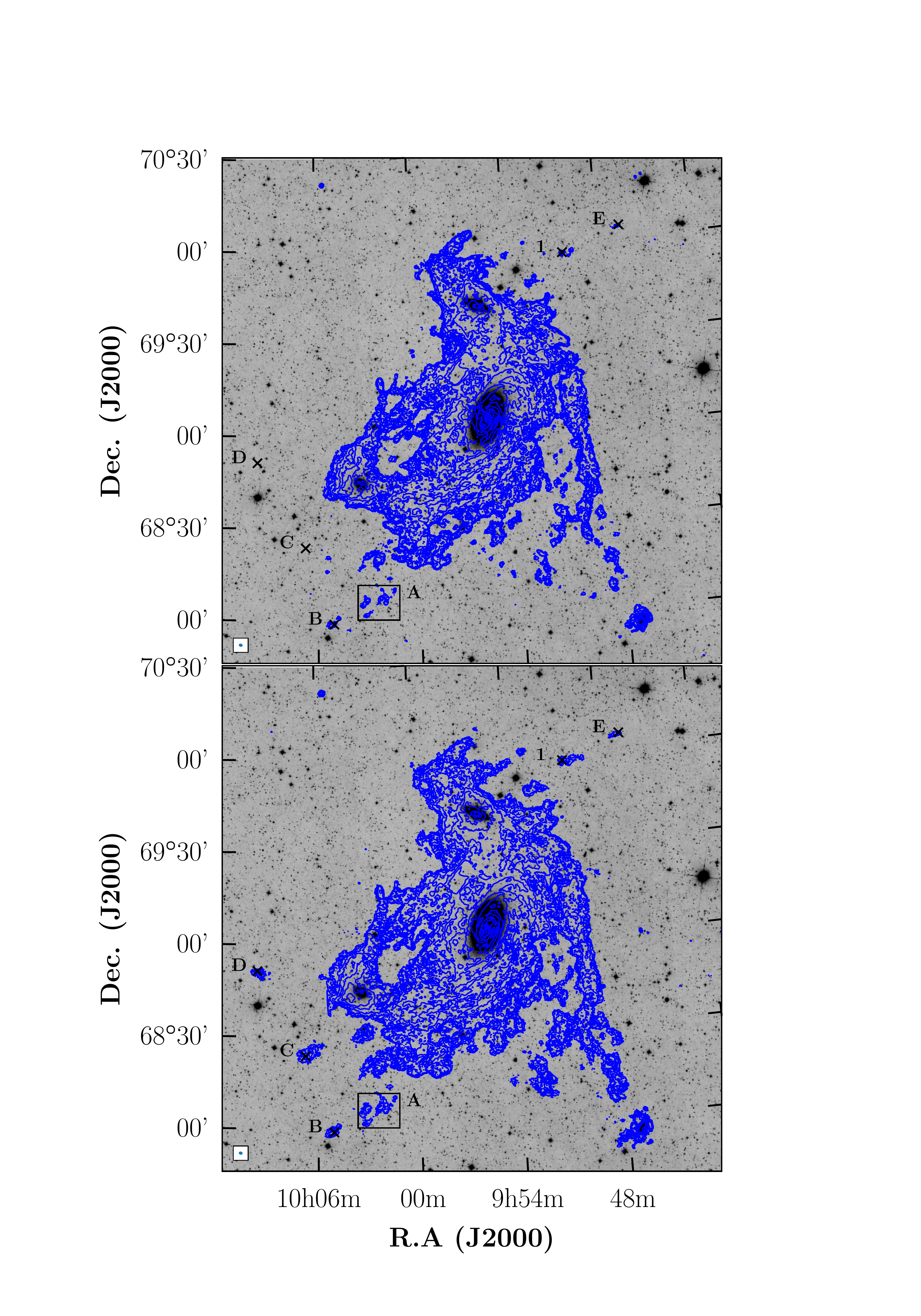}}
\vspace{-50pt}
\caption{The VLA(C+D)+GBT total \hi\ intensity maps of the M81 system. {\it Top panel}: zero-spacing correction performed using GBT data, adapted from \citet{DeBlok2018a}; {\it bottom panel}: zero-spacing correction done with EBHIS data. Contours are $2, 4, 8,\dots\times10^{19}\,\cm$.}\label{fig:vla-nhi}
\end{figure*}

In \Cref{fig:vla-nhi} we show a comparison of the re-processed VLA+EBHIS data and the GBT zero-spacing (VLA+GBT) data published in \citet{DeBlok2018a}. 
It is clear that our re-processing of the VLA data with other short baselines has improved the detection of gas with respect to \citet{DeBlok2018a}. In particular, the western arm  connected to NGC 2976 is neater, and the emission between that arm and the ``main body'' of the system is more extended.
Also, the C+D map of \citet{DeBlok2018a} revealed a few low-mass \hi\ clouds in the vicinity of the system, which they labelled as cloud ``1'' (North-West of M82), complex ``A'' and cloud ``B'' (South of NGC 3077) in their Figure 11. These clouds are found to be more extended in the new ``EBHIS-added'' mosaic, and we detect two additional clouds east of the system.
These additional clouds, which we denote clouds ``C'' and ``D'', appear at velocities of the MW emission and their association to the system is therefore uncertain. However, the peak of cloud ``D'' appears in the GBT data of \citet{Chynoweth2008} at $4.5\e{18}\,\cm$, but it remains unclear at what velocity the peak is located.
At about $17'$ (translating to a projected distance of about 18 kpc) North-West of cloud ``1'', is located a second cloud (cloud ``E'') whose peak is seen in the \citet{DeBlok2018a} data but appears more extended in the combined data. In \Cref{tb:clouds} we list the respective \hi\ masses of the different clouds, and for completeness, we also list the \hi\ masses from \citet{Chynoweth2008} and \citet{DeBlok2018a}. We note that the VLA+EBHIS \hi\ masses of the clouds ``1'' and ``B'' are intermediate between the VLA-only and the GBT respective masses. The masses measured in the combined mosaic are about 1.5 higher that those measured in the VLA-only mosaic. This confirms that the extended structures of these clouds are not just a resolution effect, but we detect low column density gas that was previously unseen in the VLA+GBT mosaic. The mass of the complex ``A'' is higher than in both the VLA-only data and the GBT data, which previously detected two clouds and a single overdensity, respectively. In fact, the GBT mass of ``A'' and ``B'' listed in \Cref{tb:clouds} are masses corresponding to the peak flux of the respective clouds in the GBT map, quoted in \citet{DeBlok2018a}. Therefore, the listed GBT mass of ``A'' does not necessarily account for the total mass of the 3 individual clouds, and may thus underestimate the total mass.
Similarly to the DRAO mosaic, the VLA+EBHIS map of NGC 2976 shows a cloud South-East of the galaxy, and likely associated the galaxy. This cloud, whose peak only appears in the VLA+GBT map, is also seen in the GBT map as a distinct object from NGC 2976.

\begin{table*}
	\begin{tabular}{l c c c c c c}
	\hline
    \hline
	Cloud & \multicolumn{2}{c}{J2000 position} & $S_\HI$ (Jy \kms) & \multicolumn{3}{c}{$M_\HI$ ($10^7\,\Mo$)}\\
	& RA & Dec & & VLA+EBHIS & VLA-only & GBT \\
	\hline
	1 & 09 50 18.3 & 69 56 26.0 & 1.68 & 0.52 & 0.32 & 1.47\\
	A$^a$ & 10 02 23.1 & 68 06 43.3 & 4.72 & 1.48 & 0.48 & 1.2\\
	B & 10 05 05.0 & 67 58 28.9 & 1.52 & 0.47 & 0.32 & 1.2\\
	C & 10 06 42.5 & 68 23 25.1 & 3.18 & 1.00 & \dots & \dots\\
	D & 10 09 34.2 & 68 51 13.5 & 1.15 & 0.35 & \dots & \dots\\
	E & 09 46 35.7 & 70 03 57.6 & 0.56 & 1.74 & 0.02 & \dots\\
	\hline
	\end{tabular}
	\caption{\hi\ mass of the clouds detected in the VLA+EBHIS mosaic. Cloud ``1'' was named so in \citet{Chynoweth2008}, and clouds ``A'' and ``B'' in \citet{DeBlok2018a}. All masses derived in this work were determined at the distance of M81. The VLA-only masses are taken from \citet{DeBlok2018a}, and the GBT masses from \citet{Chynoweth2008}. Note: $^a$: the complex A is composed of three clouds, and the quoted mass corresponds to the total mass. The listed position corresponds to the peak of the Western cloud, which is the brightest. The peaks of the Eastern and Southern clouds have coordinates (10:03:05.7, 68:05:46.9) and (10:03:14.9,68:01:07.0), respectively.}
	\label{tb:clouds}
\end{table*}
%%%%%%%%%%%%%%%%%%%%%%%%%%%%%%%%%%%%%%%%%%%%%%%%%%%%%%%%%%%%%%%%%%%%%
%%%%%%%%%%%%%%%%%%%%%% HI KINEMATICS  %%%%%%%%%%%%%%%%%%%%%%%%%%%

\section{\hi\, kinematics of the system}\label{sec:kinematics}
Besides the major galaxies M81, M82 and NGC 3077, the existence of several dwarf galaxies has been reported in the immediate vicinity of the M81 group \citep[see e.g.,][]{Borngen1982,Borngen1985,Karachentsev2000,Karachentsev2004,Chiboucas2009,Chiboucas2013}, some of which may have formed as a result of tidal stripping \citep{Makarova2002,DeMello2008,Sabbi2008}. Obviously, each of these galaxies orbits on its own plane, and are not necessarily contained in the same plane.

The major galaxies of the group are believed to have interacted closely at least once in the past \citep{Cottrell1977,VanderHulst1979,Yun1994,Yun1999,Okamoto2015}; and the hierarchical model of the $\Lambda$CDM cosmology requires that the group was formed through the infall of the individual galaxies. Using semi analytic models, \citet{Oehm2017} investigated the possible orbits of the three major galaxies and found that both or at least one of the galaxies M82 and NGC 3077 started out from far away and was not bound to the central M81 galaxy 7 Gyr ago.

\begin{figure*}
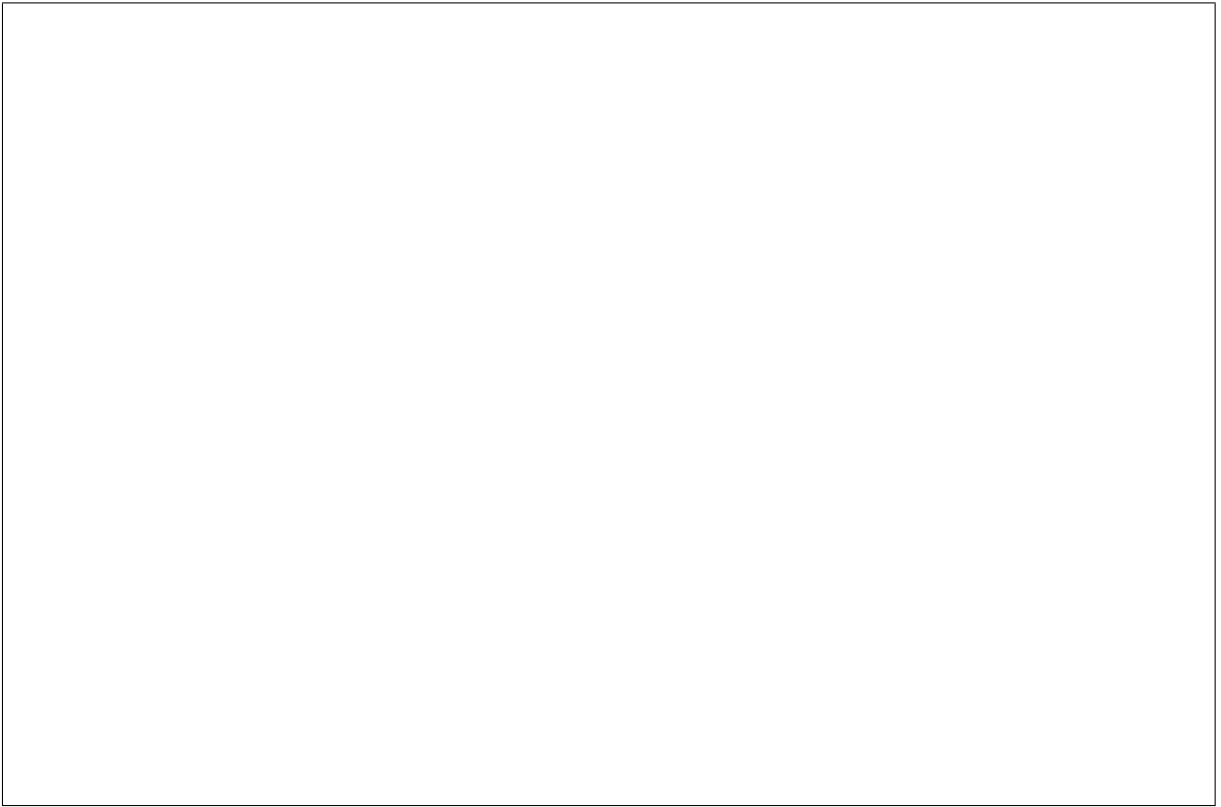

\makebox[\textwidth][c]{
\frame{
\includemedia[
activate = onclick,
width = 0.9\textwidth,
height = 0.6\textwidth,
3Dtoolbar,
3Dmenu,
3Dlights=Cube,
3Dpartsattrs=keep,
3Dviews=view.tex,
]{}{figures/composite.u3d}}}
\caption{Interactive 3D rendering of the M81 group (must be opened with Adobe Acrobat). The green contours represent the low column density envelope, the yellow contours show the intermediate column density structures, and the red solid components represent the inner regions of the major galaxies.}
\label{fig:3dviews}
\end{figure*}

In \Cref{fig:3dviews} we show an interactive 3-dimensional map \citep[inspired by Fig. 4 in][]{Hess2018} of the DRAO \hi\ datacube of the M81 group. The volume rendering was produced with the {\sc SlicerAstro} package \citep{Punzo2017}, which allows the user to manually set the flux in the data. The package contains different modules designed for viewing and analysing \hi\ line data. To optimise the volume rendering of the M81 cube, we have used a gaussian filter in the {\sc AstroSmoothing} module to smooth the cube down to a $4.8'$ resolution, and have chosen 3 separate intensity ranges to distinguish different components of the complex. These components are represented by different colours and opacities in the figure: the green transparent contours representing the \hi\ envelope of the complex, the yellow contours showing the intermediate column density component of the system containing the \hi\ northern and southern tidal bridges connecting the galaxies of the system as well as part of the western arm. This intermediate column density component also reveals a few \hi\ clouds around the system and IC 2574, and the optically dark galaxy HIJASS J1021+68. Finally, the core objects of the M81 are revealed by the high column density component shown in opaque red in the figure.

It is evident from the figure that the members of the system are not contained in the same plane, with M81 being more inclined than the plane of the envelope engulfing the individual galaxies, and the companion galaxies having different inclinations. However, despite the complexity of the geometry in this inner region of the group and the interactions between the members the common envelope seems to cut through the discs of M81 and M82 around their respective centres and presents little perturbations. Furthermore, the velocity field of the complex in \Cref{fig:vfield_drao} suggests a large-scale, ordered motion of the galaxies around the M81 galaxy, with the receding northern part that contains M82 and A0952+69 (the Arp's loop), while the southern part containing KDG61 and pointing towards BK5N and KDG64 approaches the observer. In that large-scale framework, regions at the systemic velocity of the system would contain BK3N, and the arms pointing towards NGC 3077 and Garland (East) and NGC 2976 (West) also at velocities close to systemic would act like prominent wiggles (streaming motions). It is thus tempting to fit an idealised kinematic model to the velocity field, like the usual tilted-ring model. We test that hypothesis in the following section.

\subsection{Tilted-ring model}\label{sec:tiltedring}
To measure the large-scale rotational pattern, we used the tilted ring model implemented in the \gipsy\ \citep{vanderHulst1992} task {\sc rotcur} \citep{Begeman1989}.
However, unlike for single galaxies whose rotational pattern is easily described by a circular orbit, the complex structure of the M81 group requires an extra caution. This is because the system is made of galaxies of different sizes and not necessarily orbiting in the same plane. Moreover, some of these galaxies are known to be strongly interacting, which is likely to cause non-negligible, non-circular motions that will influence the overall rotation of the large-scale system. Therefore, instead of assuming a circular orbit to describe the gas in the complex as is usually done for individual galaxies, one has to consider an additional radial, non-circular velocity component. The observed velocity is therefore the projection of the rotation and radial velocities along the line-of-sight, and can be written
\begin{equation}\label{eq:vobs}
v_{\rm obs} = v_{\rm sys} + v_{\rm rot}\,\sin{i}\,\cos{\theta} + v_{\rm rad}\,\sin{i}\,\sin{\theta},
\end{equation}
where $v_{\rm sys}$, $v_{\rm rot}$, and $v_{\rm rad}$ are respectively the systemic, rotation and radial velocities, and $i$ and $\theta$ are respectively the inclination and the azimuthal angle in the assumed plane of the complex. The latter is function of the position angle $\phi$ of the semi-major axis of the receding side of the system. 

The tilted ring model was fitted to the velocity field presented in \Cref{fig:vfield_drao}, with the optical parameters (centre position, position angle and inclination) of M81 as initial parameters. This choice of initial parameters is motivated by the fact that the velocity field is centred on the galaxy, and seems to expand out to the other two major galaxies of the system. With all parameters let free, we derived, in turn, the systemic velocity and dynamical centre of the complex. Because of the particularly complex structure of the system and to be sensitive to the local motions, we used a uniform weighting function where all points in a given ring are given the same weight.

We found that the dynamical centre is at $(\alpha, \delta)$ = (09:56:22.9, +69:03:05), with a systemic velocity of $v_{\rm sys} = -5.0$ \kms. To put this in context, the dynamical centre is located $4.5'$ east and 29 \kms\ away from M81 centre (M81's systemic velocity is $-34\pm2$ \kms, \citealt{Appleton1981}). Keeping these parameters fixed and all others left free, a third {\sc rotcur} run was used to determine the position angle and inclination of the system out to 79 kpc. The inclination is found to be increasing with the radius, while the position angle shows little variation. The velocity field of the system shows perturbations due to its complex structure, with the most noticeable being the deviation along the minor axis of M81, at the position of the dwarf galaxy HoIX. This region seems directly connected in velocity space with M82 through an arc. To exclude as much as possible these perturbations from the rotation curve, we fitted a first order polynomial to the inclination and position angle whose variations are partly caused by these perturbations. We ran {\sc rotcur} once more with these parameters fixed to the fitted values, with only the rotation velocity allowed to vary freely. The process was repeated for the approaching and the receding sides separately, and we show in the left panel of \Cref{fig:rotcurve} the variations of the position angle and inclination with the radius. The trends of the P.A for the 3 models (approaching, receding and both sides) are similar, with a highest slope noted for the approaching side. In the inner regions of the system, the approaching side shows, on average, lower values of P.A than the other two models. 
The variations of the inclinations, on the other hand, tend to be similar for the approaching side and the ``both sides'' models, while the trend of the receding side is different: while the inclination tends to increase with the radius for the approaching side and both sides, we observe a slight decrease of the inclination on the receding side, for which there is a large scatter in the outer parts. The average values of the adopted inclination and P.A for the overall model are respectively $i=62\ddeg0\pm9\ddeg1$ and $\phi=346\ddeg7\pm8\ddeg0$

\subsection{Rotation curve}\label{sec:rotcur}
\begin{figure*}
\makebox[\columnwidth][c]{
\includegraphics[width=1.15\columnwidth]{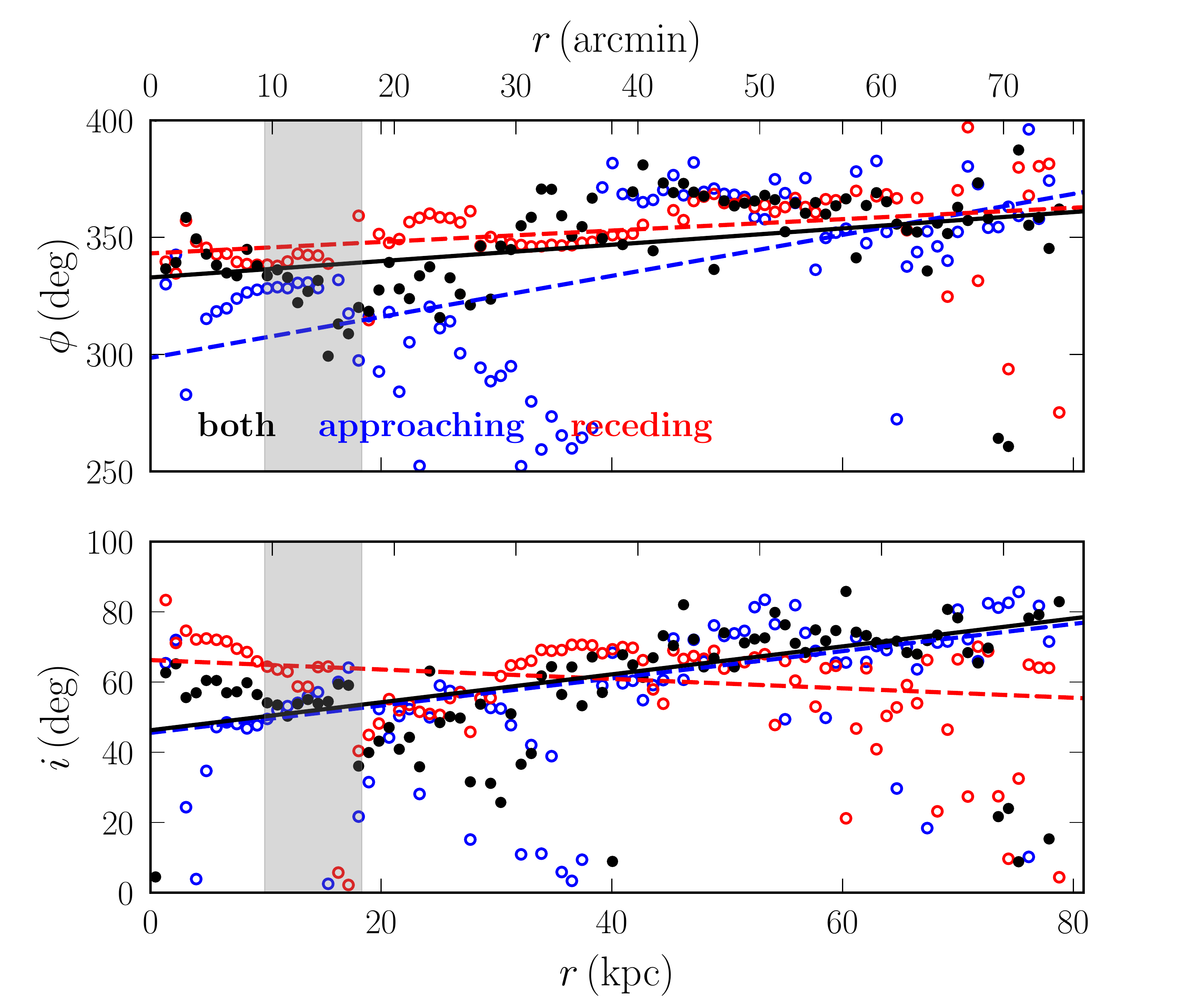}
\includegraphics[width=1.15\columnwidth]{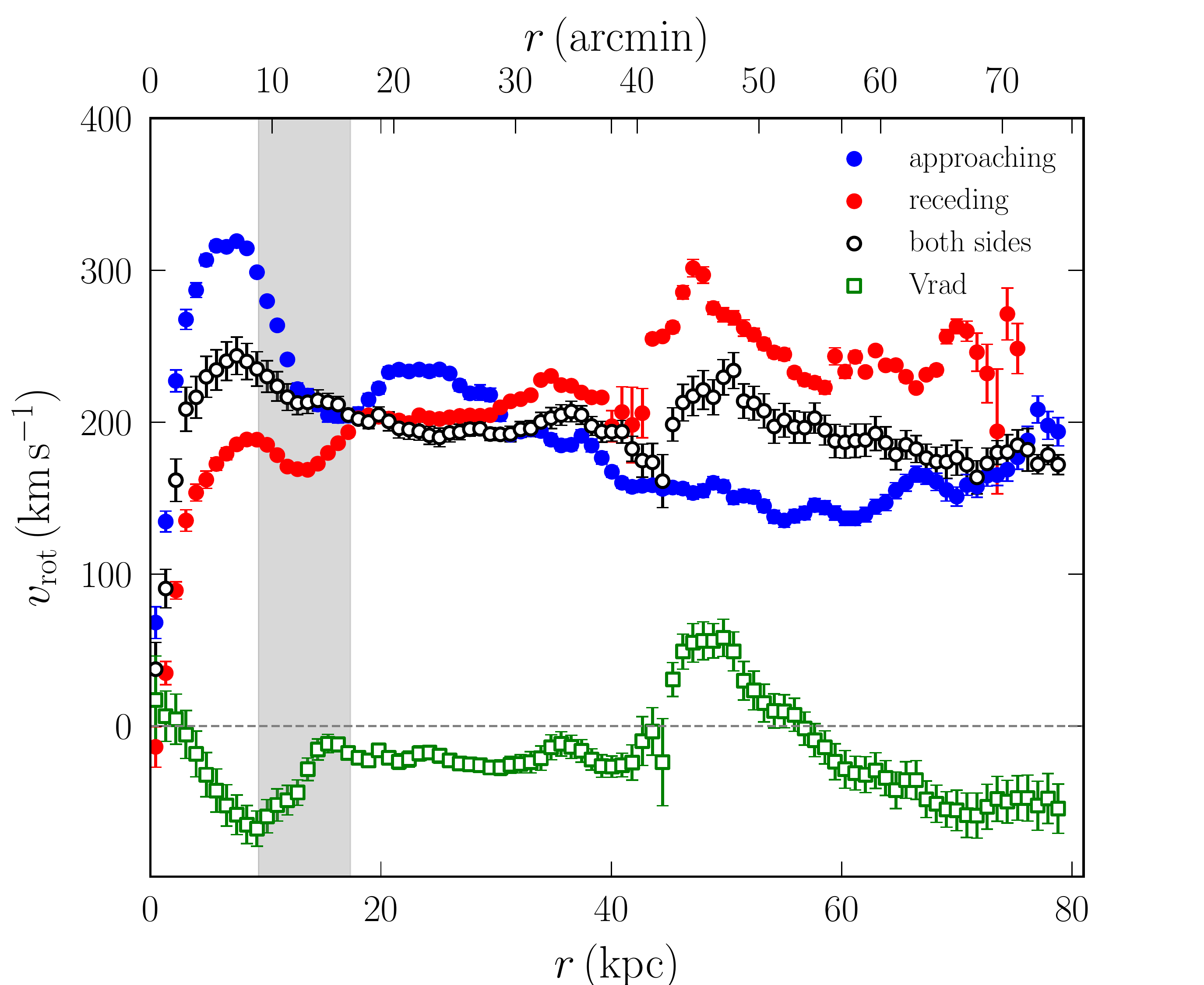}}
\vspace{-5pt}
\caption{{\it Left panel:} The variations of the position angle ({\it top}) and the inclination  ({\it bottom}) as a function of the radius in the tilted-ring model of the M81 system; {\it right panel:} Rotation curve of the M81 system derived using the tilted-ring model. The {\it green dots} show the variations of the radial velocities. The grey area in both panels corresponds to the position of HoIX.}\label{fig:rotcurve}
\end{figure*}

The rotation curve of the M81 system is presented in the right panel of \Cref{fig:rotcurve}. The uncertainties on the velocities of the average model of both sides are the quadratic sums of the uncertainties of the {\sc Rotcur} fits and half the difference between the approaching and the receding side. Also shown are the rotation curves for each of the approaching and receding sides. Not surprisingly, the approaching side is very different from the receding side, with the former peaking in the innermost region before gradually decreasing with the radius down to about 100 \kms. The receding side, on the other hand, increases slowly in the inner regions, shows signs of perturbation around the position of HoIX, and stays more or less flat in the external regions with a spike between $40'$ and $50'$, corresponding to the position of M82. This difference in the trends of the approaching and receding sides agree with the expectations, given the complex structure of the \hi\ in the complex. In fact, as noted earlier, the galaxies in the system are not contained in the same plane, and lie each at a different inclination. Each of these galaxies affects the rotation of the gas in a different way: on the receding side, the average inclination of the complex is lower than M82's inclination of $82\dg$, which tends to overestimate the rotation velocities; on the other hand, the average inclination is higher than NGC 3077's inclination of $38\dg$, which is located on the approaching side of the complex. This results in an underestimation of the velocities of the approaching side.

Despite these differences, the observation of a ``flat'' velocity curve in the outermost regions, with an amplitude comparable to that in the innermost region (the M81 disk) corroborates the visual aspect of a common rotational pattern for the interacting galaxies, that has M81 at the mass centre. 

The large-scale ordered motion is also accompanied with a significant non-circular, radial motion (green symbols). In the inner part, it is likely caused by the prominent spiral structure of M81, while in the outer regions it could reflect the respective motions of gas and galaxy companions with respect to M81.  Due to the complexity and the unknown 3D geometry of the system, it is not possible to constrain whether the radial motions are directed inwards or outwards.

\begin{figure}
\makebox[\columnwidth][c]{
\includegraphics[width=1.15\columnwidth]{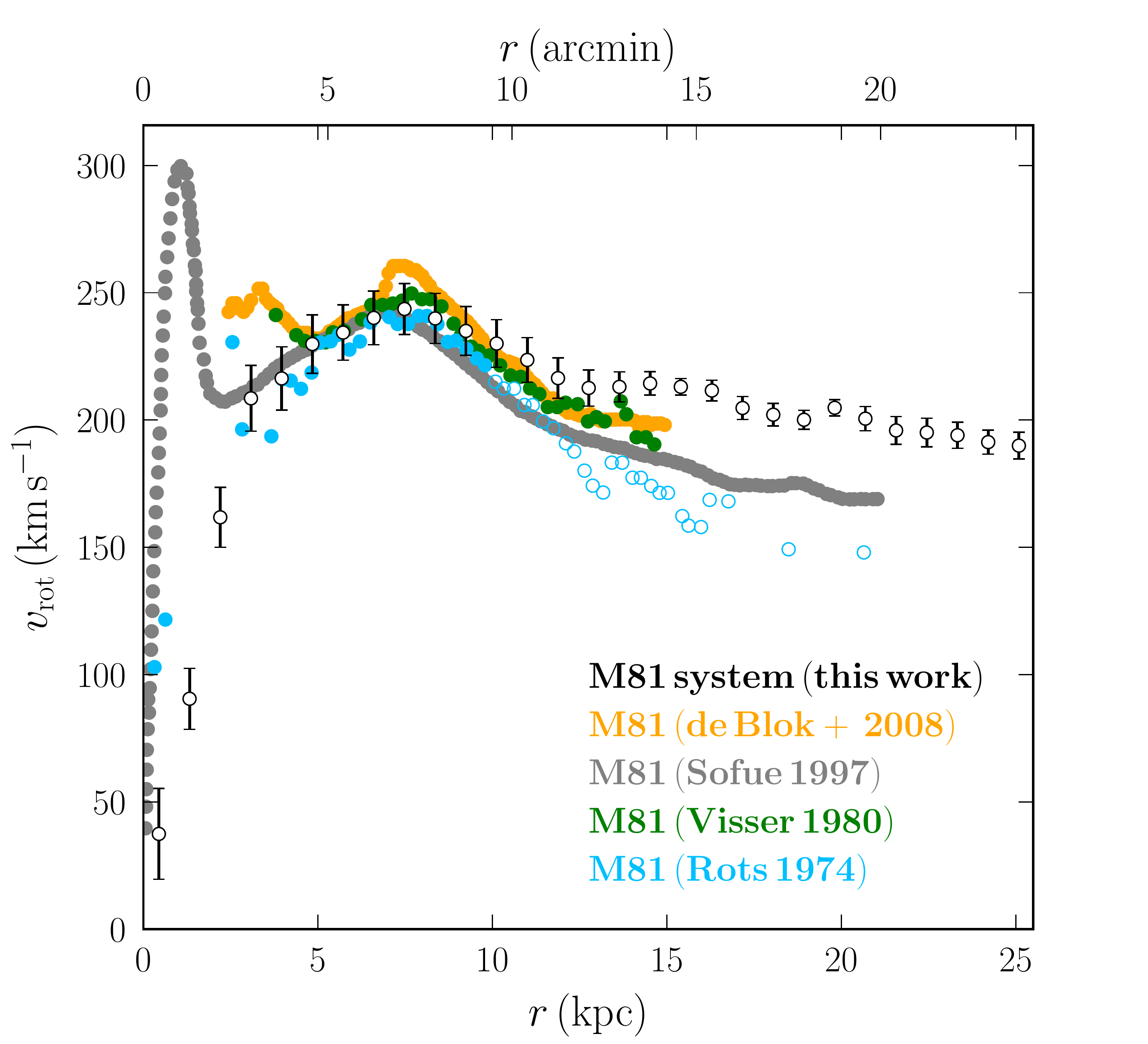}}
\vspace{-15pt}
\caption{Rotation curves of the M81 galaxy from the literature, compared to that of the M81 system in the inner 25 kpc. The {\it open circles} of \citet{Rots1974} correspond to the approaching side of the galaxy.}\label{fig:rotcurves_comp}
\end{figure}

The rotation curves of the individual galaxies M81 and M82 have been derived by a few authors (e.g., \citealt{Rots1974}, \citealt{Visser1980} and \citealt{DeBlok2008} for M81 using \hi\ data, \citealt{Sofue1997} for M81 and M82 using CO(2-1) and \hi\ data), and they revealed the complex rotations of the galaxies. The \hi\ rotation curve of M81 derived by \citet{Rots1974} extended out to $\sim20$ kpc and showed that beyond $\sim10$ kpc, the  curve of the receding side rises to high positive velocities, whereas that of the approaching side decreases monotonically. Unsurprisingly, this picture of the M81 galaxy's rotation resembles the trends that are seen in the rotation curve of the complex shown in \Cref{fig:rotcurve}, given that the galaxy constitutes the central part of the complex.
Moreover, \citet{Visser1980} and \citet{DeBlok2008} found that the rotation velocity of the galaxy is of the order of 250 \kms, and slightly decreases in the outer parts (from about 7 kpc) out to 15 kpc. This presents a picture of the galaxy whose rotation is high in the central regions, but tends to slow down as one moves to the outer regions.
On the other hand, the hybrid rotation curve of the galaxy derived in \citet{Sofue1997} showed that its rotation velocity peaks to about 300 \kms\ in the central regions, before staying slightly over 200 \kms\ out to about 10 kpc. Beyond this point, the curve decreases very slightly below 200 \kms\ out to $\gtrsim 20$ kpc. This discrepancy with the \hi\ rotation curve shows that the atomic and molecular gas rotate differently in the galaxy.
As for M82, \citet{Sofue1997} found that the galaxy's rotation curve decreases monotonically from 200 \kms\ to about 50 \kms\ in its inner $\sim4$ kpc. They later attribute this unusual shape of the galaxy's rotation curve to its interaction with M81 \citep{Sofue1998}. In \Cref{fig:rotcurves_comp} we show a comparison of the different above-mentioned rotation curves of M81 to the inner parts of system's large-scale rotation curve, where each curve was corrected to the same distance scale. Although no robust and direct comparison of the rotation of the two systems can be made because of the difference in the adopted kinematical parameters, the agreement of the different curves in the regions between $\sim5-10$ kpc shows us that the M81 system's rotation is significantly influenced by the M81's disk. The discrepancy between the system's rotation curve derived in this work and those of the literature is likely an effect of beam smearing due to the lower resolution of our data \citep[see, e.g.][]{Swaters2009,Sorgho2019}.

%%%%%%%%%%%%%%%%%%%%%%%%%%%%%%%%%%%%%%%%%%%%%%%%%%%%%%%%%%%%%%%%%%%%%
%%%%%%%%%%%%%%%%%%%%%%%%%%%% DISCUSSIONS %%%%%%%%%%%%%%%%%%%%%%%%%%%%

\section{Dynamical evolution of the group}\label{sec:discussions}
The large-scale maps derived from the present observations reveal new \hi\ features of the M81 system that help to understand the evolution of the group. The presence of gas clouds near the system, as well as the continuity of velocity in the region between the eastern side of NGC 3077 and the western side of IC 2574 suggests that there may exist more low column density gas around the system that was not detected in the present observations. This raises more questions about the formation and dynamics of the member galaxies of the group. Of course, a definite evolutionary scenario can only be obtained through numerical simulations, but the observational view offered by the present \hi\ data suggests the galaxies have formed individually at different positions, and have moved closer to each other, up to a point where they are interacting. We will discuss this scenario more in depth in this section.

The structures of the \hi\ bridges between the three members of the system suggest strong tidal interactions between the galaxies \citep[see][]{VanderHulst1979,Yun1993,Yun1994}. These interactions have caused the truncation of the M82 disk \citep{Sofue1998} and the displacement of the atomic and molecular gas of NGC 3077 from its stellar component \citep{Walter2002a}. These effects suggest that the interactions between the galaxies are violent, and they are thought to be the origin of the increased star formation rates in M81, M82 and NGC 3077 \citep{Yun1999}, in agreement with the idea proposed by \citet{Toomre1972} that such tidal disruptions could enhance starburst or AGN activity in the interacting galaxies. Such violent interactions require that at least some of the member galaxies have acquired a large momentum, which favours the idea that they have travelled a certain distance and have initiated an interaction with the other members with a high enough velocity. This is in agreement with numerical simulations of the major galaxies of the system \citep[e.g.][]{Gunn1972,Yun1999,Oehm2017}. The simulation used in \citet{Yun1999}\footnote{see an animation at \url{http://www.aoc.nrao.edu/\~myun/movie.gif}} successfully reproduces both the spatial distribution of the high column density \hi\ gas in the system, and predicts that the nearest approach of M82 and NGC 3077 to the central galaxy M81 happened respectively 220 and 280 Myr ago. This timescale is consistent with the ages of starbursts in all three galaxies, and successfully reproduces the \hi\ northern (between M82 and NGC 3077) and southern (between M81 and NGC 3077) tidal bridges revealed by the early observations of the group. However, it is now clear that there exists a tail connecting the system to NGC 2976, hinting at an interaction between the galaxy and the system. Although a further numerical analysis is required to determine the orbit of NGC 2976, the tail mapped in \hi\ shows an evidence that the galaxy has recently passed near the position of at least one of the system's members. 

The same hypothesis can be formulated, although with less confidence, for IC 2574; the galaxy may have crossed the path of at least some of the galaxies in the system. Indeed, multi-wavelength studies of IC 2574 have revealed that the galaxy has undergone two significant starburst events in the past $\sim300$ Myr \citep{Pasquali2008,Weisz2009}. Although these starbursts were attributed to stellar feedback in the galaxy, there is a possibility that its recent interaction with the system may have somehow contributed to enhance its star formation activity. 

Also, the existence of the optically dark galaxy HIJASS J1021+68 and its connection with IC 2574 further supports this hypothesis of interaction of IC 2574 with the system. In fact, \citet{Boyce2001} hinted that HIJASS J1021+68 is a possible tidal dwarf galaxy in which star formation has not begun. The present observations (bottom panel of \Cref{fig:nhimap_drao}) revealed additional clouds in the neighbourhood of the  HIJASS J1021+68, and an \hi\ envelope of IC 2574 that seems to point towards the tidal dwarf. These new features suggest that HIJASS J1021+68, as well as the neighbouring clouds, may in fact be a complex of \hi\ gas that was initially in IC2574. 

While most of the observational features of the group revealed in this work corroborate existing numerical simulations, it is not clear what the role of IC 2574 is with respect to the evolution of the M81 system. There is no direct evidence of a past interaction between the galaxy and the system, and one can only speculate from the observations. Part of the scenario is supported by existing numerical simulations, but further studies are required to account for the newly discovered low column density features. In particular, the nature of the \hi\ complex west of IC 2574 (including HIJASS J1021+68) is unclear, and this region of the group will be investigated more in detail in an upcoming paper. As stated in \Cref{sec:obs}, additional DRAO observations are underway, and will be presented in subsequent papers. These observations will undoubtedly unveil more \hi\ features of the M81 group, that will help to better constrain numerical models of the dynamical evolution of the group \citep[e.g.,][]{Oehm2017}.

%%%%%%%%%%%%%%%%%%%%%%%%%%%%%%%%%%%%%%%%%%%%%%%%%%%%%%%%%%%%%%%%%%%%%
%%%%%%%%%%%%%%%%%%%%%%%%%%% CONCLUSIONS %%%%%%%%%%%%%%%%%%%%%%%%%%%%%

\section{Conclusions}\label{sec:conclusions}
We used the DRAO synthesis telescope to map a 25 squared degrees area of the \hi\ in the M81 group. The aim of the observations was to make a full census of the \hi\ in the M81 system and the eastern region extending to the ``isolated'' galaxy IC 2574. Similarly to previous \hi\ observations of the group, we found that the interacting three major galaxies of the group -- M81, M82 and NGC 3077 -- are connected through \hi\ bridges and intergalactic \hi\ clouds. We also mapped the full extent of the western \hi\ arm connecting the three galaxies to NGC 2976, the fourth galaxy of the group located at a projected distance of 87.3 kpc southwest of M81. We have mapped, to a higher resolution than the previous GBT observations of \citet{Chynoweth2008}, the gas clouds residing in the space between the M81 system and the arm. The observations revealed that what was previously thought, from the GBT observations, to be an extended cloud is in fact a complex of small clouds filling the space between the arm and the ``main body'' of the system.

The galaxy HIJASS J1021+68, previously detected with the 76m Lovell \citep{Boyce2001,Lang2003} and the 100m GBT \citep{Chynoweth2008} single-dish telescopes, was resolved in the present observations. The DRAO mosaic in the present work suggests that the connection between the galaxy and IC 2574 \citep[previously hinted by][]{Boyce2001} is through a ``filament'' of small clouds spanning the space between the two galaxies.

Reprocessing the VLA mosaic of \citet{DeBlok2018a} with the Effelsberg single-dish data in lieu of the GBT as the authors did, we have obtained a datacube less affected by artefacts, and from which the MW \hi\ emission could be subtracted. We have mapped three previously detected clouds to a larger extent, and their respective measured \hi\ masses reveal that we have recovered more \hi\ fluxes that were undetected in the VLA-only and VLA+GBT mosaics.

Accounting for the ordered motion of the \hi\ gas around the M81 galaxy, and assuming that all gas and galaxies lie in a common rotating system, we have derived the rotation curve of the system using the tilted-ring model. This revealed that the dynamical centre of the system is located at a projected distance of about 4.7 kpc east of M81, and at systemic velocity of -5.0 \kms. The average inclination and position angle of the common plane of the system gas are $i=62\ddeg0\pm9\ddeg1$ and $\phi=346\ddeg7\pm8\ddeg0$, respectively. The rotation curve of the system is observed to be roughly constant over a large range of radii,  with a prominent wiggle associated to M82, as if the outermost gas and galaxies were rotating at the same speed as the inner disk of M81.  Although the velocities of the approaching and receding sides of the system differ, the flat trend corroborates the assumption of a large-scale rotating system dominated by M81. Our idealised rotating tilted-ring model is accompanied by a non-negligible radial motion, perhaps highlighting the streaming of outer gas and galaxies with respect to M81.

As the mapping at DRAO is still being performed, the sensitivity at the vicinity of IC 2574 will be increased. That will allow us to study the \hi\ environment of the dwarf galaxy more precisely, and investigate better its relation with HIJASS J1021+68, and the possible relation with the main M81 system. Future articles from this series will particularly focus on this analysis. 

%%%%%%%%%%%%%%%%%%%%%%%%%%%%%%%%%%%%%%%%%%%%%%%%%%%%%%%%%%%%%%%%%%%%%
%%%%%%%%%%%%%%%%%%%%%%%%%% ACKNOWLEDGEMENTS %%%%%%%%%%%%%%%%%%%%%%%%%

\section*{Acknowledgments}
The authors are very grateful to the staff at DRAO, particularly to Operations Manager Dr. Andrew Gray for his flexibility and assistance in observing the many fields and gathering archived data for our work.
We thank Prof. Tom H. Jarrett for providing the {\it WISE} stellar masses of the main galaxies of M81 in \Cref{tb:opticalprops}.
We acknowledge Dr. Kelley M. Hess for her enormous contribution in making the interactive 3D map presented in \Cref{fig:3dviews}. Kelley provided us with a step-by-step guide and advice on how to improve the display of the rendering.
We also thank Prof. Jacqueline van Gorkom for the interesting discussions, which has certainly added value to the quality of the paper.
The comments of the anonymous referee were very useful and helped to improve the quality of the present work, and for that we are grateful to them.
The work of CC is based upon research supported by the South African Research Chairs Initiative (SARChI) of the Department of Science and Technology (DST), the South African Radio Astronomy Observatory (SARAO) and the National Research Foundation of South Africa (NRF). The research of AS has been supported by SARChI and SARAO fellowships. The research of LC is supported by the Comit\'{e} Mixto ESO-Chile and the DGI at University of Antofagasta.

%%%%%%%%%%%%%%%%%%%%%%%%%%%%%%%%%%%%%%%%%%%%%%%%%%%%%%%%%%%%%%%%%%%%%
%%%%%%%%%%%%%%%%%%%%%%%%%% REFERENCES %%%%%%%%%%%%%%%%%%%%%%%%%%%%%%%

\bibliographystyle{mnras}
\bibliography{library}

\begin{thebibliography}{}
\makeatletter
\relax
\def\mn@urlcharsother{\let\do\@makeother \do\$\do\&\do\#\do\^\do\_\do\%\do\~}
\def\mn@doi{\begingroup\mn@urlcharsother \@ifnextchar [ {\mn@doi@}
  {\mn@doi@[]}}
\def\mn@doi@[#1]#2{\def\@tempa{#1}\ifx\@tempa\@empty \href
  {http://dx.doi.org/#2} {doi:#2}\else \href {http://dx.doi.org/#2} {#1}\fi
  \endgroup}
\def\mn@eprint#1#2{\mn@eprint@#1:#2::\@nil}
\def\mn@eprint@arXiv#1{\href {http://arxiv.org/abs/#1} {{\tt arXiv:#1}}}
\def\mn@eprint@dblp#1{\href {http://dblp.uni-trier.de/rec/bibtex/#1.xml}
  {dblp:#1}}
\def\mn@eprint@#1:#2:#3:#4\@nil{\def\@tempa {#1}\def\@tempb {#2}\def\@tempc
  {#3}\ifx \@tempc \@empty \let \@tempc \@tempb \let \@tempb \@tempa \fi \ifx
  \@tempb \@empty \def\@tempb {arXiv}\fi \@ifundefined
  {mn@eprint@\@tempb}{\@tempb:\@tempc}{\expandafter \expandafter \csname
  mn@eprint@\@tempb\endcsname \expandafter{\@tempc}}}

\bibitem[\protect\citeauthoryear{Appleton, Davies  \& Stephenson}{Appleton
  et~al.}{1981}]{Appleton1981}
Appleton P.~N.,  Davies R.~D.,   Stephenson R.~J.,  1981, \mn@doi [MNRAS]
  {10.1093/mnras/195.2.327}, 195, 327

\bibitem[\protect\citeauthoryear{Begeman}{Begeman}{1989}]{Begeman1989}
Begeman K.~G.,  1989, A{\&}A, 223, 47

\bibitem[\protect\citeauthoryear{Blitz, Spergel, Teuben, Hartmann  \&
  Burton}{Blitz et~al.}{1999}]{Blitz1999}
Blitz L.,  Spergel D.~N.,  Teuben P.~J.,  Hartmann D.,   Burton W.~B.,  1999,
  \mn@doi [ApJ] {10.1086/306963}, 514, 818

\bibitem[\protect\citeauthoryear{B{\"{o}}rngen \& Karachentseva}{B{\"{o}}rngen
  \& Karachentseva}{1982}]{Borngen1982}
B{\"{o}}rngen F.,  Karachentseva V.~E.,  1982, \mn@doi [AN]
  {10.1002/asna.2103030303}, 303, 189

\bibitem[\protect\citeauthoryear{B{\"{o}}rngen \& Karachentseva}{B{\"{o}}rngen
  \& Karachentseva}{1985}]{Borngen1985}
B{\"{o}}rngen F.,  Karachentseva V.~E.,  1985, \mn@doi [AN]
  {10.1002/asna.2113060508}, 306, 301

\bibitem[\protect\citeauthoryear{Boyce et~al.,}{Boyce et~al.}{2001}]{Boyce2001}
Boyce P.~J.,  et~al., 2001, \mn@doi [ApJ] {10.1086/324176}, 560, L127

\bibitem[\protect\citeauthoryear{Brouillet, Baudry, Combes, Kaufman  \&
  Bash}{Brouillet et~al.}{1991}]{Brouillet1991}
Brouillet N.,  Baudry A.,  Combes F.,  Kaufman M.,   Bash F.,  1991, A{\&}A,
  242, 35

\bibitem[\protect\citeauthoryear{Cayatte, Balkowski, van Gorkom  \&
  Kotanyi}{Cayatte et~al.}{1990}]{Cayatte1990}
Cayatte V.,  Balkowski C.,  van Gorkom J.~H.,   Kotanyi C.,  1990, \mn@doi [AJ]
  {10.1086/115545}, 100, 604

\bibitem[\protect\citeauthoryear{Chandar, Tsvetanov  \& Ford}{Chandar
  et~al.}{2001}]{Chandar2001}
Chandar R.,  Tsvetanov Z.,   Ford H.~C.,  2001, \mn@doi [AJ] {10.1086/322128},
  122, 1342

\bibitem[\protect\citeauthoryear{Chemin, Carignan  \& Foster}{Chemin
  et~al.}{2009}]{Chemin2009}
Chemin L.,  Carignan C.,   Foster T.,  2009, \mn@doi [ApJ]
  {10.1088/0004-637X/705/2/1395}, 705, 1395

\bibitem[\protect\citeauthoryear{Chiboucas, Karachentsev  \& Tully}{Chiboucas
  et~al.}{2009}]{Chiboucas2009}
Chiboucas K.,  Karachentsev I.~D.,   Tully R.~B.,  2009, \mn@doi [AJ]
  {10.1088/0004-6256/137/2/3009}, 137, 3009

\bibitem[\protect\citeauthoryear{Chiboucas, Jacobs, Tully  \&
  Karachentsev}{Chiboucas et~al.}{2013}]{Chiboucas2013}
Chiboucas K.,  Jacobs B.~A.,  Tully R.~B.,   Karachentsev I.~D.,  2013, \mn@doi
  [AJ] {10.1088/0004-6256/146/5/126}, 146, 126

\bibitem[\protect\citeauthoryear{Chynoweth, Langston, Yun, Lockman, Rubin  \&
  Scoles}{Chynoweth et~al.}{2008}]{Chynoweth2008}
Chynoweth K.~M.,  Langston G.~I.,  Yun M.~S.,  Lockman F.~J.,  Rubin K. H.~R.,
   Scoles S.~A.,  2008, \mn@doi [AJ] {10.1088/0004-6256/135/6/1983}, 135, 1983

\bibitem[\protect\citeauthoryear{Chynoweth, Langston, Holley-Bockelmann  \&
  Lockman}{Chynoweth et~al.}{2009}]{Chynoweth2009}
Chynoweth K.~M.,  Langston G.~I.,  Holley-Bockelmann K.,   Lockman F.~J.,
  2009, \mn@doi [AJ] {10.1088/0004-6256/138/1/287}, 138, 287

\bibitem[\protect\citeauthoryear{Chynoweth, Langston  \&
  Holley-Bockelmann}{Chynoweth et~al.}{2011}]{Chynoweth2011}
Chynoweth K.~M.,  Langston G.~I.,   Holley-Bockelmann K.,  2011, \mn@doi [AJ]
  {10.1088/0004-6256/141/1/9}, 141, 9

\bibitem[\protect\citeauthoryear{Cottrell}{Cottrell}{1977}]{Cottrell1977}
Cottrell G.~A.,  1977, \mn@doi [MNRAS] {10.1093/mnras/178.4.577}, 178, 577

\bibitem[\protect\citeauthoryear{Davidge}{Davidge}{2009}]{Davidge2009}
Davidge T.~J.,  2009, \mn@doi [ApJ] {10.1088/0004-637X/697/2/1439}, 697, 1439

\bibitem[\protect\citeauthoryear{Dressler}{Dressler}{1980}]{Dressler1980}
Dressler A.,  1980, \mn@doi [ApJ] {10.1086/157753}, 236, 351

\bibitem[\protect\citeauthoryear{Gendre, Fenech, Beswick, Muxlow  \&
  Argo}{Gendre et~al.}{2013}]{Gendre2013}
Gendre M.~A.,  Fenech D.~M.,  Beswick R.~J.,  Muxlow T. W.~B.,   Argo M.~K.,
  2013, \mn@doi [MNRAS] {10.1093/mnras/stt231}, 431, 1107

\bibitem[\protect\citeauthoryear{Goobar et~al.,}{Goobar
  et~al.}{2014}]{Goobar2014}
Goobar A.,  et~al., 2014, \mn@doi [ApJ] {10.1088/2041-8205/784/1/L12}, 784, L12

\bibitem[\protect\citeauthoryear{Gunn \& Gott}{Gunn \& Gott}{1972}]{Gunn1972}
Gunn J.~E.,  Gott J. R.~I.,  1972, \mn@doi [ApJ] {10.1086/151605}, 176, 1

\bibitem[\protect\citeauthoryear{Hernquist}{Hernquist}{1992}]{Hernquist1992}
Hernquist L.,  1992, \mn@doi [ApJ] {10.1086/172009}, 400, 460

\bibitem[\protect\citeauthoryear{Hess et~al.,}{Hess et~al.}{2018}]{Hess2018}
Hess K.~M.,  et~al., 2018, eprint arXiv:1811.12405

\bibitem[\protect\citeauthoryear{Hibbard \& van Gorkom}{Hibbard \& van
  Gorkom}{1996}]{Hibbard1996}
Hibbard J.~E.,  van Gorkom J.~H.,  1996, \mn@doi [AJ] {10.1086/117815}, 111,
  655

\bibitem[\protect\citeauthoryear{Hopkins, Hernquist, Cox, {Di Matteo},
  Robertson  \& Springel}{Hopkins et~al.}{2006}]{Hopkins2006}
Hopkins P.~F.,  Hernquist L.,  Cox T.~J.,  {Di Matteo} T.,  Robertson B.,
  Springel V.,  2006, \mn@doi [ApJS] {10.1086/499298}, 163, 1

\bibitem[\protect\citeauthoryear{Jacobs, Rizzi, Tully, Shaya, Makarov  \&
  Makarova}{Jacobs et~al.}{2009}]{Jacobs2009}
Jacobs B.~A.,  Rizzi L.,  Tully R.~B.,  Shaya E.~J.,  Makarov D.~I.,   Makarova
  L.,  2009, \mn@doi [AJ] {10.1088/0004-6256/138/2/332}, 138, 332

\bibitem[\protect\citeauthoryear{Karachentsev et~al.,}{Karachentsev
  et~al.}{2000}]{Karachentsev2000}
Karachentsev I.~D.,  et~al., 2000, A{\&}A, 363, 117

\bibitem[\protect\citeauthoryear{Karachentsev, Karachentseva, Huchtmeier  \&
  Makarov}{Karachentsev et~al.}{2004}]{Karachentsev2004}
Karachentsev I.~D.,  Karachentseva V.~E.,  Huchtmeier W.~K.,   Makarov D.~I.,
  2004, \mn@doi [AJ] {10.1086/382905}, 127, 2031

\bibitem[\protect\citeauthoryear{Karachentseva, Karachentsev  \&
  Shcherbanovsky}{Karachentseva et~al.}{1979}]{Karachentseva1979}
Karachentseva V.~E.,  Karachentsev I.~D.,   Shcherbanovsky A.~L.,  1979, AISAO,
  11, 3

\bibitem[\protect\citeauthoryear{Kenney, Geha, J{\'{a}}chym, Crowl, Dague,
  Chung, van Gorkom  \& Vollmer}{Kenney et~al.}{2014}]{Kenney2014}
Kenney J. D.~P.,  Geha M.,  J{\'{a}}chym P.,  Crowl H.~H.,  Dague W.,  Chung
  A.,  van Gorkom J.,   Vollmer B.,  2014, \mn@doi [ApJ]
  {10.1088/0004-637X/780/2/119}, 780, 119

\bibitem[\protect\citeauthoryear{Landecker et~al.,}{Landecker
  et~al.}{2000}]{Landecker2000}
Landecker T.~L.,  et~al., 2000, \mn@doi [A{\&}AS] {10.1051/aas:2000257}, 145,
  509

\bibitem[\protect\citeauthoryear{Lang et~al.,}{Lang et~al.}{2003}]{Lang2003}
Lang R.~H.,  et~al., 2003, \mn@doi [MNRAS] {10.1046/j.1365-8711.2003.06535.x},
  342, 738

\bibitem[\protect\citeauthoryear{Makarova et~al.,}{Makarova
  et~al.}{2002}]{Makarova2002}
Makarova L.~N.,  et~al., 2002, \mn@doi [A{\&}A] {10.1051/0004-6361:20021426},
  396, 473

\bibitem[\protect\citeauthoryear{Matsushita, Kawabe, Kohno, Matsumoto, Tsuru
  \& Vila-Vilaro}{Matsushita et~al.}{2005}]{Matsushita2005}
Matsushita S.,  Kawabe R.,  Kohno K.,  Matsumoto H.,  Tsuru T.~G.,
  Vila-Vilaro B.,  2005, \mn@doi [ApJ] {10.1086/425408}, 618, 712

\bibitem[\protect\citeauthoryear{Mattila, Fraser, Smartt, Meikle,
  Romero-Canizales, Crockett  \& Stephens}{Mattila et~al.}{2013}]{Mattila2013}
Mattila S.,  Fraser M.,  Smartt S.~J.,  Meikle W. P.~S.,  Romero-Canizales C.,
  Crockett R.~M.,   Stephens A.,  2013, \mn@doi [MNRAS] {10.1093/mnras/stt202},
  431, 2050

\bibitem[\protect\citeauthoryear{Mayya, Carrasco  \& Luna}{Mayya
  et~al.}{2005}]{Mayya2005}
Mayya Y.~D.,  Carrasco L.,   Luna A.,  2005, \mn@doi [ApJ] {10.1086/432644},
  628, L33

\bibitem[\protect\citeauthoryear{Mihos \& Hernquist}{Mihos \&
  Hernquist}{1996}]{Mihos1996}
Mihos J.~C.,  Hernquist L.,  1996, \mn@doi [ApJ] {10.1086/177353}, 464, 641

\bibitem[\protect\citeauthoryear{Neininger, Guelin, Klein, Garcia-Burillo  \&
  Wielebinski}{Neininger et~al.}{1998}]{Neininger1998}
Neininger N.,  Guelin M.,  Klein U.,  Garcia-Burillo S.,   Wielebinski R.,
  1998, A{\&}A, 339, 737

\bibitem[\protect\citeauthoryear{Nilson}{Nilson}{1973}]{Nilson1973}
Nilson P.,  1973, Acta Universitatis Upsaliensis. Nova Acta Regiae Societatis
  Scientiarum Upsaliensis - Uppsala Astronomiska Observatoriums Annaler

\bibitem[\protect\citeauthoryear{Oehm, Thies  \& Kroupa}{Oehm
  et~al.}{2017}]{Oehm2017}
Oehm W.,  Thies I.,   Kroupa P.,  2017, \mn@doi [MNRAS]
  {10.1093/mnras/stw3381}, 467, 273

\bibitem[\protect\citeauthoryear{Okamoto, Arimoto, Ferguson, Bernard, Irwin,
  Yamada  \& Utsumi}{Okamoto et~al.}{2015}]{Okamoto2015}
Okamoto S.,  Arimoto N.,  Ferguson A. M.~N.,  Bernard E.~J.,  Irwin M.~J.,
  Yamada Y.,   Utsumi Y.,  2015, \mn@doi [ApJ] {10.1088/2041-8205/809/1/L1},
  809, L1

\bibitem[\protect\citeauthoryear{Pasquali et~al.,}{Pasquali
  et~al.}{2008}]{Pasquali2008}
Pasquali A.,  et~al., 2008, \mn@doi [ApJ] {10.1086/591658}, 687, 1004

\bibitem[\protect\citeauthoryear{Punzo, van~der Hulst, Roerdink, Fillion-Robin
  \& Yu}{Punzo et~al.}{2017}]{Punzo2017}
Punzo D.,  van~der Hulst J.~M.,  Roerdink J. B. T.~M.,  Fillion-Robin J.~C.,
  Yu L.,  2017, \mn@doi [A{\&}C] {10.1016/j.ascom.2017.03.004}, 19, 45

\bibitem[\protect\citeauthoryear{Rots}{Rots}{1974}]{Rots1974}
Rots A.~H.,  1974, PhD thesis, Univerity of Groningen, \url
  {http://adsabs.harvard.edu/abs/1974PhDT.......114R}

\bibitem[\protect\citeauthoryear{Sabbi, Gallagher, Smith, de Mello  \&
  Mountain}{Sabbi et~al.}{2008}]{Sabbi2008}
Sabbi E.,  Gallagher J.~S.,  Smith L.~J.,  de Mello D.~F.,   Mountain M.,
  2008, \mn@doi [ApJ] {10.1086/587548}, 676, L113

\bibitem[\protect\citeauthoryear{Sault, Teuben  \& Wright}{Sault
  et~al.}{1995}]{Sault1995}
Sault R.~J.,  Teuben P.~J.,   Wright M. C.~H.,  1995, ASPC, 77, 433

\bibitem[\protect\citeauthoryear{Serra, Jurek  \& Floer}{Serra
  et~al.}{2012}]{Serra2012a}
Serra P.,  Jurek R.,   Floer L.,  2012, \mn@doi [PASA] {10.1071/AS11065}, pp
  296--300

\bibitem[\protect\citeauthoryear{Serra et~al.,}{Serra et~al.}{2015}]{Serra2015}
Serra P.,  et~al., 2015, \mn@doi [MNRAS] {10.1093/mnras/stv1326}, 452, 2680

\bibitem[\protect\citeauthoryear{Sofue}{Sofue}{1998}]{Sofue1998}
Sofue Y.,  1998, \mn@doi [PASJ] {10.1093/pasj/50.2.227}, 50, 227

\bibitem[\protect\citeauthoryear{Sofue \& Reich}{Sofue \&
  Reich}{1979}]{Sofue1979}
Sofue Y.,  Reich W.,  1979, A{\&}AS, 38, 251

\bibitem[\protect\citeauthoryear{Sofue \& Yoshiaki}{Sofue \&
  Yoshiaki}{1997}]{Sofue1997}
Sofue Y.,  Yoshiaki 1997, \mn@doi [PASJ] {10.1093/pasj/49.1.17}, 49, 17

\bibitem[\protect\citeauthoryear{Sorgho et~al.,}{Sorgho
  et~al.}{2019}]{Sorgho2019}
Sorgho A.,  et~al., 2019, \mn@doi [MNRAS] {10.1093/mnras/sty2785}, 482, 1248

\bibitem[\protect\citeauthoryear{Swaters, Sancisi, van Albada  \& van~der
  Hulst}{Swaters et~al.}{2009}]{Swaters2009}
Swaters R.~A.,  Sancisi R.,  van Albada T.~S.,   van~der Hulst J.~M.,  2009,
  \mn@doi [A{\&}A] {10.1051/0004-6361:200810516}, 493, 871

\bibitem[\protect\citeauthoryear{Taylor et~al.,}{Taylor
  et~al.}{2003}]{Taylor2003}
Taylor A.~R.,  et~al., 2003, \mn@doi [AJ] {10.1086/375301}, 125, 3145

\bibitem[\protect\citeauthoryear{Thomasson \& Donner}{Thomasson \&
  Donner}{1993}]{Thomasson1993}
Thomasson M.,  Donner K.~J.,  1993, Astronomy and Astrophysics, Vol.272, NO.
  1/MAYI, P. 153, 1993, 272, 153

\bibitem[\protect\citeauthoryear{Toomre \& Toomre}{Toomre \&
  Toomre}{1972}]{Toomre1972}
Toomre A.,  Toomre J.,  1972, \mn@doi [ApJ] {10.1086/151823}, 178, 623

\bibitem[\protect\citeauthoryear{Visser}{Visser}{1980}]{Visser1980}
Visser H. C.~D.,  1980, A{\&}A, 88, 149

\bibitem[\protect\citeauthoryear{Vollmer, Braine  \& Pappalardo}{Vollmer
  et~al.}{2008}]{Vollmer2008b}
Vollmer B.,  Braine J.,   Pappalardo C.,  2008, A{\&}A, 464, 455

\bibitem[\protect\citeauthoryear{Walter, Weiss, Martin  \& Scoville}{Walter
  et~al.}{2002a}]{Walter2002a}
Walter F.,  Weiss A.,  Martin C.,   Scoville N.,  2002a, \mn@doi [AJ]
  {10.1086/324633}, 123, 225

\bibitem[\protect\citeauthoryear{Walter, Weiss  \& Scoville}{Walter
  et~al.}{2002b}]{Walter2002}
Walter F.,  Weiss A.,   Scoville N.,  2002b, \mn@doi [ApJ] {10.1086/345287},
  580, L21

\bibitem[\protect\citeauthoryear{Walter, Brinks, de Blok, Bigiel, Kennicutt,
  Thornley  \& Leroy}{Walter et~al.}{2008}]{Walter2008}
Walter F.,  Brinks E.,  de Blok W. J.~G.,  Bigiel F.,  Kennicutt R.~C.,
  Thornley M.~D.,   Leroy A.,  2008, \mn@doi [AJ]
  {10.1088/0004-6256/136/6/2563}, 136, 2563

\bibitem[\protect\citeauthoryear{Weisz, Skillman, Cannon, Walter, Brinks, Ott
  \& Dolphin}{Weisz et~al.}{2009}]{Weisz2009}
Weisz D.~R.,  Skillman E.~D.,  Cannon J.~M.,  Walter F.,  Brinks E.,  Ott J.,
  Dolphin A.~E.,  2009, \mn@doi [ApJ] {10.1088/0004-637X/691/1/L59}, 691, L59

\bibitem[\protect\citeauthoryear{Winkel, Kerp, Fl{\"{o}}er, Kalberla, Bekhti,
  Keller  \& Lenz}{Winkel et~al.}{2016}]{Winkel2016}
Winkel B.,  Kerp J.,  Fl{\"{o}}er L.,  Kalberla P. M.~W.,  Bekhti N.~B.,
  Keller R.,   Lenz D.,  2016, \mn@doi [A{\&}A] {10.1051/0004-6361/201527007},
  585, A41

\bibitem[\protect\citeauthoryear{Wolfe, Pisano, Lockman, McGaugh  \&
  Shaya}{Wolfe et~al.}{2013}]{Wolfe2013}
Wolfe S.~A.,  Pisano D.~J.,  Lockman F.~J.,  McGaugh S.~S.,   Shaya E.~J.,
  2013, \mn@doi [Nature] {10.1038/nature12082}, 497, 224

\bibitem[\protect\citeauthoryear{Wolfe, Lockman  \& Pisano}{Wolfe
  et~al.}{2016}]{Wolfe2016}
Wolfe S.~A.,  Lockman F.~J.,   Pisano D.~J.,  2016, \mn@doi [ApJ]
  {10.3847/0004-637X/816/2/81}, 816, 81

\bibitem[\protect\citeauthoryear{Wright et~al.,}{Wright
  et~al.}{2010}]{Wright2010}
Wright E.~L.,  et~al., 2010, \mn@doi [AJ] {10.1088/0004-6256/140/6/1868}, 140,
  1868

\bibitem[\protect\citeauthoryear{Yun}{Yun}{1999}]{Yun1999}
Yun M.~S.,  1999, IAUS, 186, 81

\bibitem[\protect\citeauthoryear{Yun, Ho  \& Lo}{Yun et~al.}{1993}]{Yun1993}
Yun M.~S.,  Ho P. T.~P.,   Lo K.~Y.,  1993, \mn@doi [ApJ] {10.1086/186901},
  411, L17

\bibitem[\protect\citeauthoryear{Yun, Ho  \& Lo}{Yun et~al.}{1994}]{Yun1994}
Yun M.~S.,  Ho P.~T.,   Lo K.~Y.,  1994, \mn@doi [Nature] {10.1038/372530a0},
  372, 530

\bibitem[\protect\citeauthoryear{de Blok, Walter, Brinks, Trachternach, Oh  \&
  Kennicutt}{de~Blok et~al.}{2008}]{DeBlok2008}
de Blok W. J.~G.,  Walter F.,  Brinks E.,  Trachternach C.,  Oh S.-H.,
  Kennicutt R.~C.,  2008, \mn@doi [AJ] {10.1088/0004-6256/136/6/2648}, 136,
  2648

\bibitem[\protect\citeauthoryear{de Blok et~al.,}{de~Blok
  et~al.}{2018}]{DeBlok2018a}
de Blok W. J.~G.,  et~al., 2018, AJ, 865, 26

\bibitem[\protect\citeauthoryear{de Mello, Smith, Sabbi, Gallagher, Mountain
  \& Harbeck}{de~Mello et~al.}{2008}]{DeMello2008}
de Mello D.~F.,  Smith L.~J.,  Sabbi E.,  Gallagher J.~S.,  Mountain M.,
  Harbeck D.~R.,  2008, \mn@doi [AJ] {10.1088/0004-6256/135/2/548}, 135, 548

\bibitem[\protect\citeauthoryear{de Vaucouleurs, de Vaucouleurs, {Corwin, H.
  G.}, Buta, Paturel  \& Fouqu{\'{e}}}{de~Vaucouleurs
  et~al.}{1991}]{DeVaucouleurs1991}
de Vaucouleurs G.,  de Vaucouleurs A.,  {Corwin, H. G.} J.,  Buta R.~J.,
  Paturel G.,   Fouqu{\'{e}} P.,  1991, {Third Reference Catalogue of Bright
  Galaxies. Volume I: Explanations and references. Volume II: Data for galaxies
  between 0h and 12h. Volume III: Data for galaxies between 12h and 24h}.
Springer, \url {http://adsabs.harvard.edu/abs/1991S{\&}T....82Q.621D}

\bibitem[\protect\citeauthoryear{van Dokkum, Franx, Fabricant, Kelson  \&
  Illingworth}{van Dokkum et~al.}{1999}]{VanDokkum1999}
van Dokkum P.~G.,  Franx M.,  Fabricant D.,  Kelson D.~D.,   Illingworth G.~D.,
   1999, \mn@doi [ApJ] {10.1086/312154}, 520, L95

\bibitem[\protect\citeauthoryear{van~der Hulst}{van~der
  Hulst}{1979}]{VanderHulst1979}
van~der Hulst J.~M.,  1979, A{\&}A, 75, 97

\bibitem[\protect\citeauthoryear{van~der Hulst, Terlouw, Begeman, Zwitser  \&
  Roelfsema}{van~der Hulst et~al.}{1992}]{vanderHulst1992}
van~der Hulst J.~M.,  Terlouw J.~P.,  Begeman K.~G.,  Zwitser W.,   Roelfsema
  P.~R.,  1992, ASPC, 25, 131

\makeatother
\end{thebibliography}

% Don't change these lines
\bsp	% typesetting comment
\label{lastpage}
\end{document}